# Solar Temperature Variations Computed from SORCE SIM Irradiances Observed During 2003-2020

Robert F. Cahalan[1] · Paulino Ajiquichí[2] · Gaspar Yatáz[2]

**Abstract**

NASA's Solar Radiation and Climate Experiment (SORCE) Spectral Irradiance Monitor (SIM) instrument produced about 17 years of daily average Spectral Solar Irradiance (*SSI*) data for wavelengths 240 nm – 2416 nm. We choose a day of minimal solar activity, 2008-08-24, during the 2008 − 2009 minimum between cycles 23 and 24, and compute the brightness temperature ($T_o$) from that day's solar spectral irradianc*e (SSI$_o$)*. We consider small variations of *T* and *SSI* about these reference values, and derive linear and quadratic analytic approximations by Taylor expansion about the reference day values. To determine approximation accuracy, we compare to exact brightness temperatures *T* computed from the Planck spectrum, by solving analytically for *T*, or equivalent root-finding in Wolfram Mathematica. We find that the linear analytic approximation overestimates, while the quadratic underestimates the exact result. This motivates search for statistical "fit" models "in between" the two analytic models, with minimum root-mean-square-error RMSE. We make this search using open-source statistical R software, determine coefficients for linear and quadratic fit models, and compare statistical with analytic RMSE's. When only linear analytic and fit models are compared, the fit model is superior at ultraviolet, visible, and near infrared wavelengths. This again holds true when comparing only quadratic models. Quadratic is superior to linear for both analytic and statistical models, and statistical fits give smallest RMSE's. Lastly, we use linear analytic and fit models to find an interpolating function in wavelength, useful in case the SIM results need adjustment to another choices of wavelengths, to compare or extend to any other instrument.

✉ R. F. Cahalan    robert.f.cahalan@nasa.gov

[1]    Goddard Space Flight Center, Greenbelt, MD 20771
[2]    Universidad del Valle de Guatemala, Ciudad de Guatemala, Zona 15, Vista Hermosa III.

## 1. Introduction

The Sun's temperature and its variations over timescales from hours to decades have been determined since 1978 from satellite measurements of associated variations in Total Solar Irradiance (*TSI*). Since the deployment of the SORCE satellite in 2003, the Sun's temperature has also been determined for a continuous range of wavelengths that span ultraviolet, visible and near infrared wavelengths, from solar spectral irradiance (*SSI*) measurements across the peak of the *SSI* distribution. These are of great interest due to the fundamental role that solar variations play in understanding variations of the Earth's climate (Harder et al., 2005; Eddy, 2009). Beyond decadal timescales, the solar irradiance is key to estimating the Sun's luminosity, and on the longest timescales it determines Earth's lifetime, since it determines when the Sun will exhaust its energy from fusion of hydrogen in the core (Bahcall, 2000).

The average radiative temperature of the Earth is determined by an approximate balance between the amount of energy it receives from the Sun, which can be calculated from the *TSI* and the Earth's albedo, and the amount of energy that Earth emits into space that depends on Earth's emissivity (Stephens et al., 2015). The Earth's albedo is the fraction of solar energy reflected back into space, which averages about 30%, the remainder being absorbed by the atmosphere and surface (Wild et al., 2013). To determine how solar variations impact Earth's atmosphere-ocean system at various heights, SSI must be monitored in addition to *TSI*.

The relationship between Earth's temperature and the variability of solar irradiance was first speculated by Herschel, and as observations have improved, so has understanding of solar variability and its contribution to climate change (Gray et al., 2010; Bahcall, 2000). The variability of both *TSI* and *SSI* occurs due to variations in magnetic fields on the solar surface, which in turn cause the appearance of sunspots and faculae (Shapiro et al, 2015). Various models attempt to predict these changes over a wide range of timescales.

Prior to SORCE, limited observations of *SSI* made studying the variability of *SSI* and solar brightness temperature difficult, but databases over several years now enable these calculations that are important for both Heliospheric and Earth sciences (Rottman et al, 2002). In this manuscript we present a study of linear and quadratic analytic and statistical approximations of the solar brightness temperature, *T*, using either a single "reference day" during solar minimum, or using statistical properties over many days in the available record. The estimation of values using linear and quadratic approximations, both analytic and statistical, are of great help in simplifying the calculations of *T*, and in interpreting its variability. In order to determine the accuracy of the approximated values, it is necessary to compare them with very nearly exact values of *T*, calculated from the monochromatic analytic *T* equation derived from the Planck distribution, or equivalently by applying root-finding techniques to the Equation that implicitly determines *T* from observed values of *SSI*.

The manuscript is structured as follows: Section 2 describes the data analyzed here, provides the online link to download it, and describes the temporal and wavelength range. Section 3 summarizes the methodology used for the analysis of the *SSI* spectral data, the calculation of exact values of solar brightness temperatures *T*, as well as linear and quadratic analytic approximations of *T*. Section 4 discusses results of exact computations of *T*, time series of observed *SSI*, and time series comparisons of exact and approximate *T* values. Section 5 concludes by summarizing the key results and suggests future directions for research related to variations in solar irradiance and brightness temperature. Finally, the manuscript contains nine appendices that derive results referenced in Sections 1 through 5. Several figures and tables are discussed throughout. Readers may interact with several of the plotted results by going to the following dashboard that was coded in Microsoft Power BI. Dashboard link: http://wayib.org/solar-temperature-variations-relative-to-a-quiet-sun-day-in-august-2008/.

## 2. Data

The *TSI* and *SSI* data were downloaded from the University of Colorado's LASP Interactive Solar Irradiance Datacenter (LISIRD), based on measurements made by instruments onboard the Solar Radiation and Climate Experiment (SORCE) satellite. The SORCE Total Irradiance Monitor (TIM) instrument provides records of Total Solar Irradiance (*TSI*), while the Spectral Irradiance Monitor (SIM) instrument provides records of the Solar Spectral Irradiance (*SSI*). Both instruments provide daily averages, with TIM beginning 2003-02-25, and SIM beginning 2003-04-14 and both ending 2020-02-25 when the SORCE instruments were passivated (i.e. turned off). We employ throughout the latest "final" data versions, v19 for *TSI*, and v27 for *SSI*, as discussed in Kopp (2020) and Harder (2020), respectively. All dates in this manuscript are given in the format YYYY-MM-DD, in accord with https://www.iau.org/static/publications/stylemanual1989.pdf .

The SIM measures *SSI* as a function of wavelength over the range from 240 nm to 2416 nm. Though measurements of *SSI* were made prior to SORCE, for example by the UARS SOLSTICE (operating during 1991-2001), SIM was the first to provide *SSI* for a continuous range of wavelengths across the peak of the solar spectrum that occurs near 500 nm, and well into the near Infrared (IR) wavelengths, with sufficient precision to determine true solar variations (see e.g. Harder, 2009; Lee *et al*., 2016).

Note that all irradiance data from SORCE, including all *TSI* and *SSI* values, are adjusted to the mean Earth-Sun distance of one astronomical unit, 1 AU (1 au). Doppler corrections are also made to remove any variations due to the satellite orbit. Absolute and relative calibrations are enabled by a variety of Laboratory measurements carried out at both University of Colorado's Laboratory and Space Physics, and at NIST facilities. Onboard instrument degradation is monitored and corrected. Our focus in this paper is on day-to-day variability at near ultraviolet, visible, and near-infrared wavelengths. For this we rely primarily on the high precision and repeatability of TIM and SIM, more than on the absolute calibration. The high quality of TIM and SIM data has been amply documented in the literature.

Due to operational difficulties encountered, particularly after 2011 as SORCE aged, there are a limited number of days where the records are given as NA (not available) or no values were recorded. These were omitted in all calculations reported here. As an example of the *SSI* records measured by the SIM instrument, the time series of the solar spectrum from 2003 to 2020 is shown in Figure 4 for a fixed wavelength, 656.20 nm, which corresponds to the hydrogen alpha ($H_\alpha$) transition in the Balmer series.

For much of the data analysis, open-source *R* and *Python* software was used, as well as commercial software including *Wolfram Mathematica*, and *Microsoft Excel*. *Mathematica* enabled precise computation of the brightness temperatures of the *SSI* data, using efficient interpolation and root-finding methods, and provided a check on exact values computed from the analytic equation for *T* derived from the Planck distribution for the spectral irradiance, shown in Appendix D.

For more details on the *TSI* and *SSI* data used here, see the "release notes" for SORCE TIM v19, and for SORCE SIM v27, available from the NASA Goddard Space Flight Center Earth Sciences Data and Information Services Center, or from the University of Colorado Laboratory for Atmospheric and Space Physics (Harder, 2020); Kopp, 2020).

## 3. Methodology

For the radiation from a blackbody, the irradiance spectrum may be computed theoretically using the Planck distribution. However, the Sun is not a perfect blackbody, due to wavelength-dependent processes in the Sun's atmosphere. Large deviations from the Planck distribution are observed, as we show below. Still it is very useful for interpreting irradiance observations to define a solar "brightness temperature," either for the *TSI*, integrating all wavelengths, or for the solar *spectral* irradiance, *SSI*, at each available wavelength. This is the temperature for which the irradiance computed from a Planck distribution coincides with the irradiance observed by an instrument outside Earth's atmosphere, for example TIM for the wavelength-integrated irradiance, the *TSI*, or SIM for the wavelength spectrum of irradiance, *SSI*.

Computation of the brightness temperature from *TSI*, $T_{eff}$, is simply a matter of explicitly solving the Stefan-Boltzmann Law for $T_{eff}$, with a result proportional to the one-quarter power of *TSI*. Appendices A, B and C discuss the importance of TSI and related quantities. Appendix D displays the Equations (D2) and (D3) which determine the value of the *spectral* brightness temperature *T* as an *explicit* function of the observed *SSI* for each fixed wavelength. Equations (D1) and (D3) also determine *T* as an *implicit* function of the observed SSI, by solving Equation (D3) for *T* as a function of *SSI* at each fixed wavelength using a root-finding procedure. We employ a root-finding algorithm developed in *Wolfram Mathematica*, using the following initial condition $T = 5770\ K$, where *T* is chosen near the effective radiative temperature computed using $TSI = 1360.8$ W/m$^2$ as provided by the SORCE TIM (Kopp and Lean, 2011). These two approaches produce the same values of *T*, referred to in this paper as the "exact" values, and each method provides a check on the other.

SORCE SIM provides a daily SSI record for each associated wavelength from 240 nm to 2416 nm, so that during the 17 years there is a large amount of data. To handle the large number of records, algorithms were developed in *R, Python* and *Mathematica*, to provide *approximate* values of *T*. These approximate alternatives allow more rapidly computed values of *T* for any date, given a fixed set of wavelengths. In this manuscript we investigate linear and quadratic analytic approximations as a function of the observed *SSI* values, derived in Appendix E. Below it will be shown that these approximations bracket the exact values, which motivates the development of linear and quadratic *fit* approximations, that minimize the root-mean-square-error (RMSE) across a large range of days, which can include all available days. These *fit* approximations are developed in Appendix (G).

For the development of the linear and quadratic analytic approximations, a Taylor expansion is used (see Appendix E). Having the derivatives of *T* with respect to the *SSI*, this expansion gives a representation of *T* in terms of polynomial functions of *SSI*. To keep the models simple, only the first and second terms of this expansion are considered.

To apply a Taylor expansion it is necessary to have a reference value around which to expand. For this, we choose the *SSI* on a single "reference" day during the 2008-2009 solar minimum of cycle 23. Namely we choose 2008-08-24, and label that day's exact values ($T_o$, $SSI_o$). With the observed value of $SSI_o$ and associated computed value of $T_o$ during solar minimum, the linear and quadratic coefficients were calculated for the analytic approximation models. The remainder of this section discusses the time series of *SSI* and estimated *T* values. The following section 4 compares the approximate values with the exact values, and also compares the *analytic* approximations with analogous *fit* approximations that use coefficients obtained by minimizing RMSE (root-mean-square-errors) over all days, and also over two selected ranges of days.

To compare the estimation with the exact value of brightness temperature, we compute difference values, and relative differences, or delta values as:

$$Delta = \frac{T_{exact} - T_{estimate}}{T_{exact}} \tag{3.1}$$

In addition to the linear and quadratic analytic approximations obtained with the Taylor expansion, a linear and quadratic fit model is developed in Appendix G, with the help of *R* statistical software. The linear and quadratic fit models have coefficients that depend on a given temporal range of available data, and not only on the chosen reference day as is the case with the analytic approximations. In section 4 we report results for the full range of available days, as well as for two subranges, those of "early" and "late" days, R1 and R2, respectively.

The comparison between the brightness temperatures calculated with the linear and fit approximations are shown in the tables, along with a comparison between the linear coefficients. In order to make the computations very explicit, in Appendix H, an example of the calculation of the brightness temperature is given for the linear and quadratic *analytic* approximation methods, as well as for the linear and quadratic *fit* approximation methods, for a randomly selected day.

In Appendix I, a method of rapid interpolation is given for the linear analytic and fit coefficients, valid over a broad range of wavelengths that satisfies $400\ nm \leq \lambda \leq 1800\ nm$.

## 4. Results

Before considering the *temporal* variations of *SSI* observed by the SORCE TIM and SIM instruments over the 17 years, 2003-2020, we first consider the *wavelength* variations of *SSI* on our chosen "reference day" 2008-08-24. Figure 1 shows this $SSI_o$ wavelength dependence observed on the reference day, in green, and for comparison the Planck irradiance distributions computed for temperatures $T = 4500\ K, 5770\ K, 6500\ K$ using Equation (D1), in blue, tan, and red, respectively. The lower and upper Planck temperatures are seen to give computed *SSI* values that bracket the observations of $SSI_0$ for this wavelength range, while the computed *SSI* for the intermediate 5770 K (tan) approximately follows the observed $SSI_0$ (green). Although the observed value coincides with the computed 5770 K Planck value for a few wavelengths only, otherwise the observed values of $SSI_0$ lie above or below the Planck curve. The measured value of *TSI* (historically "solar constant") by the Total Irradiance Monitor (TIM) instrument on the reference day is $TSI_o = 1360.4704\ W/m^2$ and is associated with an effective radiative temperature of $T_o = 5771.2685\ K$, close to $T = 5770\ K$, used in computing the intermediate tan curve in Figure 1 (Kopp and Lean, 2011).

Figure 2 is a zoom of Figure 1 for wavelength range 240 nm to 660 nm. The apparently irregular bumps in this plot, and in Figure 1, are due to well-known Frauenhofer lines in the solar spectrum, smoothed to the SIM instrument's bandpass, which varies from about 1 nm width near wavelength 200 nm, up to almost 30 nm near 1000 nm, then decreasing slightly (Harder *et al.*, 2005). The width of a typical atomic Frauenhofer line is of order 1 Angstrom, or 0.1 nm, so the observed bumps are smoothed clusters of several nearby lines. A few of the contributing atomic lines are indicated in the labels on the vertical dashed lines. For example, the green dashed line near 430 nm, is labelled CaFeg to indicate that lines of calcium, iron, and oxygen (g-band) are all included within the plotted bump in the green line. Effects of ionization thresholds are also seen, such as just above the CaII H and K lines near 400 nm, which has photon energies near 3.1 eV.

*TSI* provides key observational data about the Sun and is needed to compute the Sun's luminosity and lifetime (see Appendix B). *TSI* is not a solar *constant*, as had been assumed prior to the satellite era. Its value varies due to turbulent magnetic processes on the Sun. *TSI* variations amount to about 0.1% (1000 ppm) of the mean value over the four solar cycles so far observed by satellite (cycles 21 through 24), since 1978. The average solar luminosity, and thus the *TSI*, is determined by nuclear processes in the Sun's core. These change over a much longer timescale than the solar cycle, up to billions of years, as nuclear processes transform hydrogen into helium. The present value of *TSI*, and thus solar luminosity can provide a good estimate of the Sun's lifetime, and thus the time that the Sun's nuclear fuel will eventually run out. Such calculations are shown in Appendix B, where it is shown that the current best *TSI* value at solar minimum, $1360.80 \pm 0.50\ W/m^2$ (Kopp and Lean, 2011), gives the overall lifetime of the Sun as approximately $10.70$ billion years. The current estimated age of the Sun, and of our solar system, is about equal to Earth's estimated age of 4.54 billion years (± 50 million years). So this leaves about 6.2 billion years, more or less, before the Sun will expand into a Red Giant, leaving a white dwarf star behind.

The importance of *TSI* in climatic variability has been mentioned, for example in computing Earth's global average effective radiative temperature. Appendix C estimates the effective temperature of the Earth as $255.48\ K$, using the *TSI* on the reference day, and Earth's average albedo of $0.29$ (Stephens et al., 2015).

*TSI* is the integral of *SSI* over all wavelengths, and *SSI* in turn determines the solar spectral *brightness temperature T* at each wavelength. Determining *T* as well as *SSI* is useful in understanding the physical and chemical processes that take place on the Sun. For example, Figure 3.A, a plot of the brightness temperature, $T_o$, on the reference day, shows a broad peak above 1600 nm. This is associated with transitions in Hydrogen ions $H^-$ (1 proton + 2 electrons). Photons with a wavelength $\lambda < 1644$ nm are dominated by the $H^-$ bound free-transitions, while photons with $\lambda > 1644\ nm$ are absorbed and re-emitted in $H^-$ free-free transitions (Wildt, 1939). The $H^-$ ion is the major source of optical opacity in the Sun's atmosphere, and thus the main source of visible light for the Sun and similar stars.

Now we consider the *temporal* variations of *SSI*. Figure 4 shows the time series of the irradiance corresponding to a fixed wavelength, in particular for $H_\alpha$ ($H_\alpha$ wavelength = 656.2 nm), the longest wavelength in Hydrogen's *Balmer Series*. The variability of the *SSI* can be seen, with the deepest minimum occurring early in the record, during Oct-Nov 2003. The spike that goes below 1.523 W/m²/nm is associated with the **Halloween solar storms**, a series of solar flares and coronal mass ejections that occurred from mid-October to early November 2003, peaking around October 28–29. See for example
 https://en.wikipedia.org/wiki/Halloween_solar_storms,_2003 .

This occurred during the declining phase of solar cycle 23. On the slower year-to-year timescale, the Sun's activity declines into the much quieter period of the *solar minimum* during 2008-2009 (Kopp, 2016). The solar minimum implies about a 0.1% decrease in solar energy that arrives on Earth, causing the Earth's temperature to decrease slightly (Gray, 2010). After this solar minimum, solar activity increases again, as cycle 24 sunspots and other solar activity increase in intensity into a *solar maximum* in 2014-2015, before declining again, into a quieter minimum period of 2019-2020.

As can be seen in Figure 5, the brightness temperature time series for $H_\alpha$ is also similar to the temporal variability of the *SSI* for $H_\alpha$. It is evident that they are in phase. As *SSI* data is extended beyond the end of SIM, by TSIS-1 and successor missions, the solar cycles will become more evident, as happened with *TSI* (Solanki et al 2013). It is important to emphasize that the spectral brightness temperatures are wavelength dependent radiative temperatures of the Sun, the temperatures at which the *SSI* data measured by the satellite coincides with what is obtained using the Planck distribution (Trishchenko, 2005).

Figure 6 shows a plot of the linear analytic approximation of brightness temperature compared with the exact value, the value obtained by Equation (D2), or the root-finding solution of Equations (D1) and (D3). The linear analytic approximation is given by neglecting the quadratic term in Equation (E18), taking as reference the date during solar minimum, 2008-08-24. Figure 6 shows that this approximation closely overlays the exact.

To more clearly see the difference between the exact and the linear analytic approximation, Figure 7A shows the difference, exact – approximation, in units of mK = $10^{-3}$ K, and Figure 7B the delta, difference/exact (Equation 3.1) in parts per million (ppm). The negative differences in Figures 7A and 7B show that the linear analytic approximation **overestimates** the exact value of the brightness temperature. The root-mean-square-error (RMSE) is 0.00041244545, very small, which explains why such differences are not evident in Figure 6. A significant increase in variability is seen in 2011 and afterwards, so Figure 7A also displays the RMSE for both the earlier, quieter, period, as well as the later, noisier period. Some of this increased noise is due to solar cycle 24, but some is likely also due to the aging of the satellite and the SIM instrument.

Figure 8 shows the plot obtained using the *quadratic* analytic approximation given by Equation (E18), with (D3), together with the exact values calculated from Equation (D2) or from root-finding with (D1) and (D3). This looks nearly identical to the analogous Figure 6 for the linear analytic approximation. However, in the plot analogous to Figure 7, we plot in Figure 9 the difference between the exact and the *quadratic* analytic approximation, and here the results are quite different from the linear case. Figure 9A shows the difference, exact – approximation in units of μK = $10^{-6}$ K, and Figure 9B the delta, difference/exact in parts per million (ppm) in the *quadratic* case. The *positive* differences in Figures 9A and 9B show that the quadratic approximation **underestimates** the exact value of the brightness temperature, though are much closer than the linear, with RMSE reduced to 0.0000003428, more than 1000x smaller than the linear case in Figure 7, and Table 2 shows that the Mean Error (Bias) is also more than 1000x smaller than the linear. Comparing Figures 7 and 9 (and Table 2) indicates that the opposite signs of the bias suggests there may be a better approximation that lies "in between" the linear and quadratic approximations. Below we will show that the "fit" approximations do typically provide such improvements.

Figure 9A also shows that, in accord with intuition, the decrease in RMSE indicates that the approximate value is better the more terms are considered in the Taylor expansion. The improvement to from RMSE 7A to RMSE 9A removes the most significant figures in RMSE 7A, suggesting a rapidly converging series. This indicates that finding an improved "in between" *fit* approximation will be a challenge, as the quadratic *analytic* approximation is excellent.

Table 2 drives home this last point, comparing the RMSE for linear and quadratic *analytic* models, with the RMSE for the linear and quadratic *fit* models for the same $H_\alpha$ wavelength used in Figures 7 and 9. Indeed, though the linear *fit* model RMSE is about 2.85x **smaller** than the linear *analytic* RMSE, the *quadratic fit* model RMSE is 2.81x times **smaller again** than the 1000 x smaller q*uadratic analytic* RMSE. So, at the $H_\alpha$ wavelength, the quadratic *fit* model is more precise even than the very precise quadratic *analytic* model.

Tables 1, 3 and 4 extend Table 2 to wavelengths 285.5 nm, 855.93 nm, and 1547.09 nm, respectively. As noted for $H_\alpha$, at these near-ultraviolet and near-infrared wavelengths, the *linear fit* model also has **smaller** RMSE than the *linear analytic*. Also, if we compare the two *quadratic* models, then *again* for 255.5 nm, 855.3 nm, and 1547.09 nm, the *quadratic analytic* model wins, and for 255.5 and 855.3 it is by an even larger factor than it does for $H_\alpha$, by factors 10.34 and 7.78, respectively, while for 1547.09 nm the quadratic fit model wins over the quadratic analytic by a factor 2.00. If we take these 4 wavelengths as representative, then, the *quadratic fit* model is preferred, and nearly reproduces the exact values, despite the high precision of the quadratic analytic model.

Some applications may not require such high precision. If we choose to restrict ourselves to *linear* models the *fit* model is still preferred, though it is a close call at 1547.09 nm, where the linear analytic model RMSE is 1.05 x larger than the linear fit, so has only a 5% edge. At that wavelength the linear analytic may be sufficient, and indeed an analytic approach has some advantages. For example, it may be optimized for a particular ranges of dates of particular interest, and the single coefficient interpreted as a "linear sensitivity" of temperature to irradiance at this wavelength.

Note that the SIM instrument registers a higher variability of spectral irradiance for shorter wavelengths 285.5 nm and 355.93 nm. This occurs because the more energetic photons (according to the Planck-Einstein relationship E = hc/λ) allow for more transition and ionization processes than at near infrared wavelengths, such as those shown, 855.3 nm and 1547.09 nm.

Continuing with the plans for simplifying the calculations of the brightness temperature, which is the central objective of the manuscript, Figure 11 shows the plots of the quotients of the linear analytic coefficients for certain wavelengths. Looking at the behavior of the curve of the quotients $a'$, a polynomial interpolation was obtained as discussed in Appendix I. This provides a simple mathematical expression useful in calculating the linear coefficients for any wavelength in the range from 400 nm to 1800 nm. With this, calculating the brightness temperature becomes simpler and faster than Equation (E18) with (D3), and valid for interpolating to wavelength bins not aligned with wavelength bins measured by the SIM instrument.

To compare the linear analytic and linear fit models, Figures 12 and 13 show the differences between the coefficients of the linear analytic approximation model, Equation (E10) and (E18) omitting the quadratic term, or (G4), and that of the linear fit model, Equation (G1). Note that the fit coefficient $a$ in Equation (G1) is computed using R software, and depends on the range of days supplied. This can range over the full set of days available from SORCE SIM (17 years of daily data). For comparison we also compute $a$R1 over the set of days in the first half of the data, that have the smaller or RSME values shown in Figure 7a, as well as $a$R2 over the late day range, with larger RMSE. In short, early and late year ranges are R1=2003-2010, and R2=2011-2020. All three ranges, overall, R1 and R2 are shown in Figures 12 and 13. In the figures we can see that the values obtained with Equations (G1) and (G4) (with (E10)) do not vary much for wavelengths less than 1400 nm and greater than 400 nm, therefore the brightness temperature values that are calculated in that range of wavelengths also do not differ much, using the linear analytic and linear fit models. Note that $a$R1 and $a$R2 values lie on either side of the overall difference value of $a$, which in every case lies in between, for each wavelength.

## 5. Summary and Conclusions

Our results and conclusions may be summarized as follows: (a) The linear and quadratic analytic approximation models, Equations (E18), with Equation (E10) for the linear term, and (E17) for the quadratic term, and (E3) to compute $B$ from *SSI*, simplify calculations of solar brightness temperature $T$ on any chosen day for a fixed wavelength, with $B$ or *SSI* as a single variable. (b) The **linear** analytic approximation **overestimates** the exact values of $T$, while the **quadratic** analytic approximation **underestimates** the exact values, but has much smaller RMSE (rms error) than the linear. (c) By using the full dataset to find coefficients that minimize the RMSE we find linear and quadratic "fit" approximations that lie closer to the exact values for representative wavelengths, as can be seen by the "fit" RMSE values in Tables 1 to 4, being smaller than the corresponding analytic RMSE's, i.e. (fit RMSE)/(analytic RMSE) < 1 for both linear and quadratic cases, for near-ultraviolet, visible, and near-infrared wavelengths. (d) For wavelengths in between the tabulated ones, Equations (I1) and (I2) provide a smooth interpolating polynomial function of wavelength, simpler and faster to apply than Equation (E10) in the analytic case, or the R software in the fit case, and accurate for any wavelength within a broad range across the peak of the *SSI*, extending into near infrared wavelengths that are of particular importance in modeling Earth's climate.

The statistical measure used to understand differences between values calculated by the linear and quadratic analytic approximation models with the exact values of $T$ obtained from Equation (E2) (or root-finding in Mathematica software), is the RMSE (root mean square error). Figures 7.A and 9.A and Table 2 show that for the $H_\alpha$ wavelength the RMSE for the linear ($0.0004124455\ K$) and quadratic ($0.0000003428\ K$) analytic approximation models are small, and therefore the deviations between the estimated and exact values are small. Table 1 shows that for wavelength 285.5 nm the RMSEs for both analytic models remain small, though larger than for $H_\alpha$. For both these wavelengths, the quadratic analytic model is superior to the linear analytic model. Tables 3 and 4 shows that for the longer near infrared wavelengths 855.93 nm and 1547.09 nm this pattern continues, with the quadratic analytic model being superior to the linear analytic. The fact that at all 4 wavelengths the quadratic analytic RMSE is smaller than the linear analytic RMSE suggests that further terms in the Taylor expansion may converge towards the exact over the full wavelength range. However, we do not have a proof of convergence. Even if the series does converge, there is only a suggestion, not a guarantee, that it will converge to the exact value given by Equation (E2).

A comparison of the linear analytic coefficient (Equation (D13) or (G4)) with the coefficient of the linear least squares *fit* of the data performed with the statistical packages of *R* software are shown in Figures 11 and 12. The linear fit model shows the line that best represents the entire data set, whereas the linear analytic approximation model has its maximum accuracy on the chosen reference day. Gaps in the data, the primary one being that which occurs from July 20, 2013 to March 12, 2014 (Harder et. al, 2019) have a direct influence on the coefficient of linear fit, because the solar spectrum measurement instruments SIM A and SIM B showed significant differences from the spectrum measured at the beginning of 2011, as can be seen for example in Figure 1 of the manuscript of Harder et. al, 2019.

Despite the good quality of the two analytic approximations, we find that the two fit models provide better "in between" approximations. **The most accurate of the four approximations considered here is the quadratic fit model.** We have seen that the brightness temperatures that it produces are in most cases indistinguishable from the exact temperatures that are found as roots of the Equation that defines the brightness temperature, $SSI = \alpha_s B(T)$, where $B$ is the Planck distribution, and $\alpha_s$ the solid angle subtended by the Sun at the mean Earth distance.

There will soon be new opportunities to apply and extend this study. Both TIM and SIM instruments are now acquiring daily data onboard the International Space Station. The new record, begun in 2018-03-14, had sufficient overlap with SORCE to enable the prior dataset to be adjusted to match TSIS-1 (https://lasp.colorado.edu/lisird/data/sorce_sim_tav_l3b/ ). Currently TSIS-1 extends to 2021-07-20 and continues to be extended. TSIS-1 will be succeeded by TSIS-2, which is expected to continue the record beyond the peak of solar cycle 25. We look forward to testing and applying the approximations studied here to future solar cycle data, to enable improved understanding of the Sun's temperature variations.

## Appendices

## Appendix A. Total Solar Irradiance and Sun's effective temperature

The Sun is not a blackbody, since the brightness temperature varies significantly with wavelength, as shown in Figure 1. However, we can define an "effective" radiative temperature $T_{eff}$, using the blackbody formula, with Stefan-Boltzmann constant $\sigma = 5.670374 * 10^{-8}$ W/m²/K⁴ as follows

$$TSI = \alpha * \sigma T_{eff}^4 \qquad (A1)$$

Here *TSI* is the total solar irradiance (historically "solar constant"), while $\alpha$ is the ratio between the total area of the Sun with radius $R_s = 6.957 * 10^8$ m divided by the area of a sphere centered on the Sun with radius equal to one astronomical unit AU = 149,597,870,700.0 m, so that

$$\alpha = \frac{4\pi R_s^2}{4\pi (AU)^2} = \left(\frac{R_s}{AU}\right)^2 = 2.16268 * 10^{-5} \qquad (A2)$$

The energy flow emitted by the Sun decreases as it diverges from the Sun's photosphere, becoming isotropically decreasing as 1/distance². *TSI* and *SSI* values measured by satellites like SORCE are adjusted to the mean Earth-Sun distance of one AU, thus removing variations due to the satellite orbit. Earth receives a small fraction of the energy emitted by the Sun and recorded by satellites, and that fraction will be considered in the following.

## Appendix B. Solar luminosity and the Sun's lifetime

Questions of how the Sun shines, and how old it is, have been objects of interest since ancient times, but it was not until the scientific revolution that there was opportunity to give definitive answers, first from classical physics, then using ideas from relativity, quantum mechanics and nuclear physics. With the development of modern theories, the answer became well understood (Bethe, 1939; Bahcall, 2000; Adelberg *et al.*, 2010).

The solar luminosity, *L*, is the total solar power, the total radiative energy emitted from the Sun per second, isotropically in all directions. The best current estimate of *L* relies on the measurements of *TSI*, which is the solar power per *m²* at the mean Earth-Sun distance of one *AU*. To obtain *L* from *TSI*, multiply by the total number of square meters on a sphere with radius equal to the Earth-Sun distance, so using the TIM value from Kopp and Lean, 2011 gives

$$L = TSI * (4\pi * AU^2) = \left(1360.8 \frac{W}{m^2}\right) * (4\pi * AU^2) = 3.82696 * 10^{26} \ W \qquad (B1)$$

The energy produced by nuclear reactions in the Sun's core is determined using Einstein's $E=mc^2$, where m is the mass loss in the primary reaction, which in the Sun is conversion of four H atoms into one He, as explained in 1939 by Hans Bethe in his classic 1939 paper "Energy production in stars," for which he won the Nobel Prize (Bethe, 1939).

Assuming the Sun's luminosity is approximately constant over the Sun's lifetime, then

$$L * \text{Lifetime} = \text{Total Joules} = M_s c^2 * 0.00723 * 0.1 // (31{,}556{,}952) \qquad (B2)$$

where the Sun's mass $M_s = 1.9885 \times 10^{30}$ kg, the speed of light in vacuum $c = 299{,}792{,}458$ m/s is, the mass loss fraction in the 4H–>He process 0.00723 is, the core fraction in which nuclear reactions are self-sustaining is 0.1, and the number of seconds per year is 31,556,952. Dividing Equation (B2) by Equation (B1), the lifetime of the Sun can now be computed, with the result

$$Lifetime \approx \frac{Total\ Joules}{L} = 10.70 * 10^9 \ \text{years} \qquad (B3)$$

If the constant value of *L* is replaced by a linearly increasing *L*, while the Sun is also assumed to be about halfway through its lifetime, then the above estimate is not significantly altered, since a dimmer younger Sun is compensated by a brighter older Sun.

**Appendix C. Earth's temperature from TSI**

Earth intercepts a small fraction of the solar energy, casting a small shadow on the sphere of area $4\pi * AU^2$. That absorbed energy fraction is determined by the product of the *TSI*, the Earth's cross-section ($\pi * R_E^2$), and the Earth's absorptivity, or its albedo $\alpha = 1 -$ absorptivity. The absorbed fraction determines Earth's global mean temperature (North et al, 1981; Gray et al, 2010). Earth's temperature then determines the total thermal energy that Earth emits back into space. The **balance** between the absorbed solar energy, and the emitted thermal energy, determines Earth's effective radiative temperature, $T_E$ This condition of radiative equilibrium at the top of Earth's atmosphere is expressed as

$$TSI*(1-\alpha)*(\pi * R_E^2) = \sigma T_E^4 * (4\pi * R_E^2) \tag{C1}$$

Dividing through by Earth's surface area gives the global average energy emitted and absorbed in the form

$$\sigma T_E^4 = \left(\frac{TSI}{4}\right) * (1-\alpha) = \epsilon * (1-\alpha) \tag{C2}$$

where $\epsilon = TSI/4$, and $\alpha = Earth\ albedo = 0.29$. (Note **albedo** symbol $\alpha$ used in Equation (C2) is **not** the $\alpha$ of Equation (A2)!) Now, knowing that the energy absorbed and radiated by the Earth are equal, in thermal equilibrium, the effective temperature of the Earth can be calculated as

$$T_E = \left(\frac{E_{abs}}{\sigma}\right)^{0.25} \tag{C3}$$

TSI impacts the average and variability of Earth's temperature and, of course, its variations have impacted climate for millions of years (Solanki et al 2013; Kopp and Lean, 2011). TSI variations can be understood as a combined impact of variations in sunspots, and faculae, as well as variations occurring over the entire Sun. Models based on these have been key tools in studies of Earth's climate (e.g. Kopp and Lean, 2011; Foukal and Lean, 1985).

**Appendix D. "Exact" Solar Brightness Temperatures**

As already mentioned in previous paragraphs, the Sun is not a pure blackbody. The *SSI* (solar spectral irradiance) has evident deviations from a pure Planck distribution, due to atomic absorption and ionization processes in the solar atmosphere. (See Figures 1 and 2.) An especially helpful way to study these deviations is by transforming *SSI* at each wavelength $\lambda$ into solar brightness temperature *T*. To do this, at each fixed wavelength, we solve the Planck distribution for *T*. That is, we solve

$$B(\lambda, T) = \frac{k_1}{\lambda^5 * \left(\exp\left(\frac{k_2}{\lambda T}\right) - 1\right)} \tag{D1}$$

where $k_1 = 10^{20} c_1 = 1.19268 * 10^{20}\ W * m^2/Sr$ and $k_2 = 10^7 c_2 = 1.43877 * 10^7\ K * m$ are constants, and the units of *B* are $W/m^2/nm$. Solving for *T* gives the following, which we term the "exact" solar brightness temperature:

$$T(\lambda, B) = \frac{k_2}{\lambda * \ln\left(\frac{k_1}{\lambda^5 B} + 1\right)} \tag{D2}$$

To obtain the solar spectral irradiance *SSI* from the Planck distribution *B* requires an integral over the solid angle of the Sun at the Earth's mean orbital distance. This gives

$$SSI = \alpha_s B \tag{D3}$$

where the value of $\alpha_s = \pi * \alpha$, so from Equation (A2) we have $\alpha_s = 6.79426 * 10^{-5}$. Note the wavelength $\lambda$ is kept fixed, and for each wavelength there is a corresponding brightness temperature *T*, determined by the value of temperature for which the satellite's *SSI* observation coincides with the Planck distribution for that $\lambda$ and *T*. Equivalently to Equation (D2), to solve Equation (D3) in *Mathematica* software, we use the initial condition $T = 5770\ K$ and $\alpha_s = \pi * (R_s/AU)^2$, and apply the function FindRoot to Equation (D3), which gives the same values of *T* as the explicit "exact" Equation (D2).

The next appendix shows how to approximately calculate the brightness temperatures $T$ for any fixed wavelength, having only the observed *SSI* values (or equivalently $B$) as a variable, because all other parameters are defined on a single "reference day" so do not vary from day-to-day. It is important to remember that the wavelength is fixed, and consequently, the parameters $T_o, SSI_o, (dT/dSSI)_o$ and higher derivatives (evaluated on the reference day) vary with wavelength. For the SIM data used in this manuscript to produce the plots, the wavelengths range from 240 nm to 2416 nm. For each wavelength in this range, there is a set of parameters that can be used to determine a time series of brightness temperatures $T$ for all other days in the date range [2003-04-14 to 2020-02-26].

**Appendix E. Analytic Approximations for Brightness Temperature**

This Appendix derives two simple analytic representations of the daily brightness temperatures that take advantage of the fact that, at a given wavelength, the *SSI* values are very nearly equal from day-to-day, and typically vary by less than 1%. The analytic approximations express the daily temperature values on any given day, *T*, at each fixed wavelength, by a Taylor expansion of the exact value of *T* as an analytic function of *SSI*, as given in Equation (D2). The expansion is about the value of *SSI* and *T* on a given "reference" day ($T_o$, $SSI_o$), as follows

$$T = T_0 + \left(\frac{dT}{dSSI}\right)_o (SSI - SSI_o) + \frac{1}{2}\left(\frac{d^2T}{dSSI^2}\right)_o (SSI - SSI_o)^2 + \ldots \quad \text{(E1)}$$

We focus on the "linear approximation" that keeps just the first derivative, and then the "quadratic approximation" that keeps the first two derivatives. Higher order terms will be neglected, except in discussion of convergence. Since *SSI* is directly proportional to $B$ by a constant rescaling, as given in Equation (D3), we may write (E1) as

$$T = T_o + \left(\frac{dT}{dB}\right)_o (B - B_o) + \frac{1}{2}\left(\frac{d^2T}{dB^2}\right)_o (B - B_o)^2 + \ldots \quad \text{(E2)}$$

In order to compute the first and second derivatives via the chain rule, we introduce two new variables, $y$ and $z$, as follows. Let

$$y = \ln(z) = \ln\left[\frac{k_1}{\lambda^5 B} + 1\right] \quad \text{(E3)}$$

so that

$$z = e^y \quad \text{(E4)}$$

and

$$z - 1 = \frac{k_1}{\lambda^5 B} \quad \text{(E5)}$$

Then

$$\frac{dy}{dz} = \frac{1}{z} = \frac{1}{e^y} \quad \text{(E6)}$$

and

$$\frac{dz}{dB} = \frac{-k_1}{\lambda^5 B^2} = \frac{-\lambda^5}{k_1} * (e^y - 1)^2 \quad \text{(E7)}$$

Therefore

$$\frac{dy}{dz} * \frac{dz}{dB} = \frac{-\lambda^5}{k_1} * \frac{(e^y - 1)^2}{e^y} \quad \text{(E8)}$$

Equations (D2) and (E3) imply

$$T = \frac{k_2}{\lambda y} = T(y(z(B))) \quad \text{(E9)}$$

We may compute the derivative of Equation (E9) using the chain rule, employing Equation (E8), to obtain

$$\frac{dT}{dB} \equiv T^{(1)} = \frac{dT}{dy} * \left[\frac{dy}{dz} * \frac{dz}{dB}\right] = \frac{-k_2}{\lambda y^2} * \left[\frac{-\lambda^5}{k_1} * \frac{(e^y-1)^2}{e^y}\right] = \frac{k_2 \lambda^4}{k_1 y^2} * \left[\frac{(e^y-1)^2}{e^y}\right] \quad (E10)$$

To evaluate (E10) on the reference day, we set $B = B_o$ equal to the value on that day, compute $y = y_o$ from Equation (E3), and substitute that into Equation (E10). In order to compute the second derivative, we note that Equation (E10) is already in the form analogous to (E9), namely

$$T^{(1)} = T^{(1)}(y(z(B))) \quad (E11)$$

Therefore as in computing Equation (E10) we take the derivative of (E11) using the chain rule, (E8) and (E10) to obtain

$$T^{(2)} = \frac{dT^{(1)}}{dy} * \left[\frac{dy}{dz} * \frac{dz}{dB}\right] = \frac{k_2 \lambda^4}{k_1} * \left[\frac{d}{dy}\left(\frac{1}{y^2}\right) * \frac{(e^y-1)^2}{e^y} + \frac{1}{y^2}\frac{d}{dy}\left(\frac{(e^y-1)^2}{e^y}\right)\right] * \left[\frac{-\lambda^5}{k_1} * \frac{(e^y-1)^2}{e^y}\right] \quad (E12)$$

On the right side we applied the product rule to compute $dT^{(1)}/dy$ from Equation (E10), giving the two terms in the left square brackets, and used Equation (E8) to substitute into the right square brackets. Evaluating the first term in the left bracket of Equation (E12) allows us to factor out $1/y^2$ from both terms. We also combine the rightmost constant $-\lambda^5/k_1$ with the leftmost constant $k_2\lambda^4/k_1$ to yield the following

$$T^{(2)} = \frac{-k_2 \lambda^9}{k_1^2} * \left(\frac{1}{y^2}\right) * \left[\left(\frac{-2}{y}\right) * \frac{(e^y-1)^2}{e^y} + \frac{d}{dz}\left(\frac{(z-1)^2}{z}\right)\frac{dz}{dy}\right] * \left[\frac{(e^y-1)^2}{e^y}\right] \quad (E13)$$

We apply the product rule to the remaining derivative in the second term in the left-hand brackets, and use Equation (E4) which implies $dz/dy = z$, to give

$$T^{(2)} = \frac{-k_2 \lambda^9}{k_1^2} * \left(\frac{1}{y^2}\right) * \left[\left(\frac{-2}{y}\right) * \frac{(e^y-1)^2}{e^y} + \left(\frac{2(z-1)}{z} - \frac{(z-1)^2}{z^2}\right) * z\right] * \left[\frac{(e^y-1)^2}{e^y}\right] \quad (E14)$$

In the second term within the left square brackets, we distribute the z, then factor out $\frac{(z-1)^2}{z}$ to yield

$$T^{(2)} = \frac{-k_2 \lambda^9}{k_1^2} * \left(\frac{1}{y^2}\right) * \left[\left(\frac{-2}{y}\right) * \frac{(e^y-1)^2}{e^y} + \left(\frac{2z}{z-1} - 1\right) * \frac{(z-1)^2}{z}\right] * \left[\frac{(e^y-1)^2}{e^y}\right] \quad (E15)$$

We substitute $\frac{(z-1)^2}{z} = \frac{(e^y-1)^2}{e^y}$ and factor that from both bracketed terms in Equation (E15), and combine the remaining terms with their common denominator to give

$$T^{(2)} = \frac{-k_2 \lambda^9}{k_1^2} * \left(\frac{1}{y^2}\right) * \left[\frac{-2(z-1)+y(z+1)}{y(z-1)}\right] * \left[\left(\frac{(e^y-1)^2}{e^y}\right)^2\right] \quad (E16)$$

Finally, using $z = e^y$ and simplifying gives the final result

$$T^{(2)} = \frac{-k_2 \lambda^9}{k_1^2} * [2 + y + e^y(y-2)] * \left[\left(\frac{e^y-1}{y}\right)^3 / e^{2y}\right] \quad (E17)$$

To evaluate Equation (E17) on the reference day, just as for Equation (E10), we set $B = B_o$, equal to the value observed on that day, substitute that into Equation (E3) to compute $y = y_o$, and substitute the value of $y = y_o$ into Equation (E17). Substituting these first and second derivatives of *T* evaluated on the reference day into Equation (E2), and neglecting all higher-order terms, we obtain the *quadratic analytic approximation* given by

$$T = T_o + \left(\frac{dT}{dB}\right)_o (B - B_o) + \frac{1}{2}\left(\frac{d^2T}{dB^2}\right)_o (B - B_o)^2 \quad (E18)$$

where the linear term is computed using (E10), and the quadratic term is computed using (E17). Omitting the quadratic term in (E18) gives the *linear analytic approximation*.

**Appendix F. The Sun's effective temperature**

Here we derive a linear approximation for the "effective" temperature $T_{eff}$, Equation (F5), associated with the *total solar irradiance, TSI*. This is a simpler case than for *SSI*, since for *TSI*, the Stefan-Boltzmann equation makes the *exact* $T_{eff}$ a simple analytic function of *TSI,* given in Equation (F1). As mentioned before, the Sun is not a blackbody, but we can calculate its associated effective temperature by using the Stefan-Boltzmann Equation and using *TSI* (total solar irradiance) measured directly by satellites above the atmosphere, by solving Equations (A1) and (A2) to obtain

$$T_{eff} = \left(\frac{TSI}{\alpha\sigma}\right)^{0.25} \tag{F1}$$

where $\sigma = 5.670374 \times 10^{-8}$ W/m²/K⁴ is the Stefan-Boltzmann constant, and from Equation (A2) $\alpha = 2.16268 * 10^{-5}$. Taking the derivative of (F1) gives an expression for the change in effective temperature with a change in *TSI* as follows

$$\frac{dT_{eff}}{dTSI} = \frac{d}{dTSI}\left(\left(\frac{TSI}{\alpha\sigma}\right)\right)^{0.25} \tag{F3}$$

Simplifying (F3) and evaluating on the reference day gives

$$\left(\frac{dT_{eff}}{dTSI}\right)_o = \frac{1}{4} * \left(\frac{TSI_0}{\alpha\sigma}\right)^{0.25} * \frac{1}{TSI_o} = \frac{1}{4} * \frac{(T_{eff})_o}{TSI_o} \tag{F4}$$

Substituting into Equation (F4) the values on the reference day, 2008-08-24, which are $(TSI_o, (T_{eff})_o) = (1360.4704 \ W/m^2, 5771.2685 \ K)$, we obtain

$$\left(\frac{dT_{eff}}{dTSI}\right)_o = 1.06053 \frac{K}{W/m^2} \tag{F5}$$

Since *TSI* typically varies by about 0.1% or less, Equation (F5) is quite accurate for most days. An extreme case is the "Halloween" event of 2003-10-29, when a large sunspot grouping dropped the temperature about 3.6670 K below the reference day $T_{eff}$ on 2008-08-24. Equation (F5) estimates a 3.6605 K decrease from the reference day, a 0.0065 K underestimate, which is 0.1773% of the drop, or 0.0001 % of $(T_{eff})_o = 5771.2685 \ K$. Substituting the 2003 "Halloween" values of $T_{eff}$ and *TSI* into Equation (F4) gives instead of (F5) the coefficient 1.06255. This day of *minimum $T_{eff}$* is also the day of *maximum* coefficient of sensitivity over the full 17-year SORCE SIM record and is 19% larger than the coefficient on the reference day, shown in Equation (F5).

Conversely, the *minimum* coefficient over the 17 years occurs on the day of *maximum $T_{eff}$*, which is 5773.1820 K, which occurred on 2015-02-26. That minimum coefficient is 1.05947, 10% less than the reference day's ratio in (F5). The average coefficient over all days is 1.06029. The fact that this average value is 2% *less* than the reference day's implies that the linear approximation in Equation (F5) typically slightly *overestimates* the changes in $T_{eff}$. The same is true for *SSI*: the *linear analytic approximation* of brightness temperature *T*, obtained by dropping the quadratic term in Equation (D18), also has a positive mean error, or bias, as shown for four representative wavelengths in Tables 1-4. Those tables also show that inclusion of the quadratic term in (D18) largely removes this positive bias, leaving a very small mean error, and small RMSE, as discussed in more detail in the text.

In principle, there are two ways to determine the total solar irradiance (*TSI*). The first is by using the SORCE TIM instrument to obtain a direct measurement. The second way is to use the *SSI* measured by the SIM instrument and integrate as wide a range of wavelengths as possible. As expected, there is a shortfall in the value computed by integrating the SIM data compared to what is measured by TIM, mainly due to missing energy above the longest wavelengths measured by SIM, approximately 2400 nm. This TIM-SIM difference is shown in Figure 2 of the manuscript by Harder et. al 2019, and amounts to 146.128 $W/m^2$. This must be subtracted from the value measured by TIM, or added to the integrated SIM value, for comparisons to be made between TIM and SIM. In this paper we focus on *SSI*, though both *SSI* and *TSI* must be considered in the study of Earth's climatic variations.

**Appendix G. Linear and Quadratic fit model**

Comparing Figures 7 and 9 shows that the linear analytic approximation, using just the first two terms in (E18), *overestimates* the exact *T*, given in (D2), while the quadratic approximation, using all three terms in (E18), though closer to the exact, slightly *underestimates*. To consider a possible "in between" approximation, this appendix introduces linear and quadratic "fit" models. These statistical "fit" models calculate the brightness temperature as a function of wavelength using *R* software. In the linear and quadratic *analytic* approximation models discussed in earlier appendices, estimates of solar brightness temperatures *T* are made based on the measured solar spectrum of the chosen reference day, and the exact brightness temperatures computed for that day, which occurs during a time of minimum solar activity. By contrast, the fit models we discuss below take into consideration the statistical properties of the full set of daily data over the 17 years of the SORCE mission.

A statistical model that provides a least square fit to the solar spectral irradiance (*SSI*) data obtained in the *R* software with linear regression may be written as

$$T = a * SSI + b \tag{G1}$$

where $a, b$ are constants for a specific wavelength and *T* is the brightness temperature. In the same way, *R* software may compute a least squares *quadratic* fit of the form

$$T = C * SSI^2 + A * SSI + B. \tag{G2}$$

These two fit models express linear and quadratic dependences, respectively, between *SSI* and *T*. We obtain, using code developed in open-source *R* software, simple models that best fit the data, for which mean square error is minimized.

We rewrite the analytic Equation (E1) (or the equivalent (E2)), up to the linear term, in this way

$$T = T_0 + \left(\frac{dT}{dSSI}\right)_0 * SSI - \left(\frac{dT}{dSSI}\right)_0 * SSI_0 \tag{G3}$$

$$T = \left(\frac{dT}{dSSI}\right)_0 * SSI + T_0 - \left(\frac{dT}{dSSI}\right)_0 * SSI_0$$

$$T = a' * SSI + b', \tag{G4}$$

which has a similar form to Equation (G1). This will allow us to make a comparison between values of the constants that appear in Tables 7 and 9. The linear analytic coefficient, which appears in Table 6, and $SSI_o$ are constant, because they are evaluated for the data of the reference day that appears in Table 5. It is important to mention that the constant $a'$ defined in this part is the same constant given in the linear analytic term in Equation (E1).

We follow a similar procedure for the quadratic analytic model, rewriting Equation (E1) to obtain,

$$T = C' * SSI^2 + A' * SSI + B', \tag{G5}$$

which is a mathematical expression similar to Equation (G2), where the values of the constants are given as:

$$C' = \frac{1}{2}\left(\frac{d^2T}{dSSI^2}\right)_0$$

$$A' = \left(\frac{dT}{dSSI}\right)_0 - \left(\frac{d^2T}{dSSI^2}\right)_0 * SSI_0,$$

$$B' = T_0 - \left(\frac{dT}{dSSI}\right)_0 * SSI_0 + \frac{1}{2}\left(\frac{d^2T}{dSSI^2}\right)_0 * SSI_0^2.$$

The values of constants $A', B', C'$ in the analytic model, and $A, B, C$ in the fit model, are shown in Tables 8 and 10, along with values of *T* obtained with the quadratic analytic model and the quadratic fit model for certain wavelengths.

As mentioned, the linear fit is obtained with regression techniques in the R software and for this, all the available spectral irradiance data are used for a fixed wavelength and therefore, if there is a change in the range of data, the linear fit changes because an analysis is done on all the data. In this way, a partition of the available data (2003 to 2020) into an "early period" designated R1 (2003 to 2010), and a "late period" designated R2 (2011 to 2020) was made to make a comparison between the linear coefficients shows in Tables 11 and 12.

### Appendix H. Example calculations of Brightness Temperature using the analytic and fit models.

This appendix illustrates the *T* approximations by considering an example of a randomly chosen day. For this example, results from the linear and quadratic *analytic* models are compared with the results of applying the linear and quadratic *fit* models for the randomly chosen day. In Tables 1-4, the RMSE (root-mean-square-error) and the ME (mean error, or bias), computed over all the available days in the SIM v27 record, are shown for all 4 models, linear and quadratic, analytic and fit.

To better explain how the analytic and fit models work, consider the following example for the wavelength of $\lambda = 656.20$ nm ($H_\alpha$). As mentioned in this manuscript, the data of one particular day during solar minimum is taken as a reference, in this case 2008-08-24. On that "reference day", for the $H_\alpha$ wavelength, the solar spectral irradiance (*SSI*) according to the SIM instrument was, $SSI_o = 1.526558$ Wm$^{-2}$(nm)$^{-1}$, which is the value registered in the data downloaded in LISIRD, in version 27. Using Equation (D2), checked with *Mathematica* root-finding software applied to Equation (D3), it was calculated from this reference day irradiance that the exact brightness temperature is $T_o = 5772.410671\ K$. Note: We use high precision here in order to illustrate the very small RMSEs.

Next, using Equation (D3), divide *SSI$_o$* by $\alpha_s$ to compute $B_o = 22{,}468.34828$. Then using Equation (E3) we compute $y_o = 3.79838$. Substituting $y = y_o$, along with the $H_\alpha$ wavelength, and the two radiation constants $k_1$ and $k_2$, into Equation (E10) we obtain

$$\left(\frac{dT}{dSSI}\right)_0 = 973.20427 \text{ K} * \frac{\text{nm}}{\text{W/m}^2} \tag{H1}$$

Similarly, using the same $B_o$ but varying wavelength, and so varying $y_o$, Table 6 shows the linear analytic model coefficients for the wavelengths of 285.50 nm, 656.20 nm, 855.93 nm and 1547.09 nm. The linear analytic approximation is used to estimate the value of *T* for some other day, knowing the *SSI* of that day. If we choose a random day, for example 2011-10-10, the value of *SSI* of that day (see Table 5) for the wavelength of $H_\alpha$ is $SSI = 1.527622$ Wm$^{-2}$(nm)$^{-1}$. Using Equation (E18), without the quadratic term, then yields

$$T = 5772.410671 \text{ K} + 973.20427 * (1.527622 - 1.526558)\text{ K}$$

$$T = 5773.44616 \text{ K} \tag{H2}$$

This is the linear analytic approximation for the brightness temperature for the "example" date 2011-10-10. It is close, but slightly larger than, the value computed by the "exact" Equation (D2) (or root finding in *Mathematica* software), $T = 5773.44598$ K (see Table 5). The error of root-finding is very small compared to either the analytic or statistical estimates, for the relatively smooth functions involved here, so in this manuscript both the result of using Equation (D2) and the root finding result are referred to as the "exact" value.

If we consider the root mean square error (RMSE) in Table 2, the value obtained with the linear approximation, $5773.44616\ K \pm 0.00041$ K, agrees very well with the exact value, since it is within the range of values. The exact value is obtained by applying Equation (D2). The approximate result of (H2) is obtained when using the Equation (G4) with the values of the constants in Table 7.

For the quadratic analytic approximation, we calculate the brightness temperature using Equation (E18), including the quadratic coefficient from Equations (E17), with (D3) and (E3), to give

$$\left(\frac{d^2T}{dSSI^2}\right)_o = -323.64399 \text{ K} * \left(\frac{\text{nm}}{\text{W/m}^2}\right)^2 \tag{H3}$$

Similarly, the quadratic coefficients for the other wavelengths are in Table 6, but rounded to three places. Therefore, the value of the $H_\alpha$ temperature obtained by including the quadratic term in (H2) is 5773.44597715 K, which when rounded to 5 places right of the decimal, agrees spot on with the exact. If we consider the RMSE, rounded to eight places, the range of the brightness temperature is $5773.44597715 \pm 0.00000034$ K which also includes the exact value, and is much closer to the exact than is the linear.

The above was for wavelength 656.20 nm. Table 5 shows the *SSI* and *T* values for this same wavelength, as well as for wavelengths 285.50 nm, 855.93 nm and 1547.09 nm, and also the *SSI* and exact *T* values on 2011-10-10. Tables 7 and 8 shows estimated results for 2011-10-10, using the analytic Equations (G4) and (G5), respectively.

Finally, in Tables 9 and 10 the parameters of linear (G1) and quadratic (G2) fit models are shown, and the value of the estimated brightness temperature on 2011-10-10. Comparing these results with the values of the RMSE and ME listed in Tables 1, 2, 3 and 4, the results are shown to be in excellent agreement with the exact values obtained from Equation (D2) or the *Mathematica* root-finding method.

### Appendix I: Temperature sensitivity ratios and rapid interpolation

This final appendix provides a rapid method of interpolation between the measured wavelengths.

The ratio of a small change of the Sun's effective temperature divided by the associated change of the TSI is 1.06053 K/(W/m$^2$), as given by Equation (F5). The analogous spectral relationship is the linear coefficient $\boldsymbol{a'}$ of the linear analytic approximation, the ratio of the change of spectral brightness temperature divided by the associated change in the SSI, the solar spectral irradiance, from the linear term in Equation (E18), using (E10) and (D3). To match the units of this TSI sensitivity ratio, it is appropriate to divide $\boldsymbol{a'}$ by the wavelength $\lambda$. This allows determination of the wavelength for which the ratio $\boldsymbol{a'}/\lambda$ is closest to the TSI value 1.06053 K/(W/m$^2$). The spectral values are tabulated in Table 13, which shows a minimum value of $\frac{a'}{\lambda} = 1.2280$, which occurs near 486.3 nm, whereas otherwise $\frac{a'}{\lambda} > 1.2280 > 1.06053$ K/(W/m$^2$) = *TSI* sensitivity.

For interpolation between measured wavelengths, it is useful to obtain a simple analytic mathematical expression for the brightness temperature sensitivity ratio $\boldsymbol{a'}/\lambda$. An interpolation function for the ratio $\frac{a'}{\lambda}$ may be expressed as

$$\frac{a'}{\lambda} = 6.043791 \times 10^{-6} \lambda^2 - 6.076933 \times 10^{-3} \lambda + 2.851032 \qquad (I1)$$

where $\lambda$ is any wavelength that satisfies $400\ nm \leq \lambda \leq 1800\ nm$. With the previous expression one can calculate $\boldsymbol{a'}$ for any wavelength $\lambda$ within this range, and then use the *SSI* value on any day to compute the associated brightness temperature *T*, using the linear analytic approximation, which can be written

$$T = \boldsymbol{a'} * (SSI - SSI_o) - T_o \qquad (I2)$$

Equation (I2) represents a method of calculating the brightness temperature which is simpler and faster than the linear term in Equation (E18), and valid for any wavelength within a broad range.

**Acknowledgements** We are grateful to Jerald Harder for discussions and comments on an earlier version, and to Stephane Beland for assistance obtaining the appropriate data from the LISARD system at University of Colorado's Laboratory for Atmospheric and Space Physics (LASP), and to all those at LASP who designed and built the SORCE instruments and data system, and managed SORCE throughout the 17 years. We thank NASA Goddard Space Flight Center, and the nonprofit CHEARS for sponsoring our research.

## Figure Captions

**Figure 1.** Solar Spectral Irradiance (*SSI*) vs wavelength for reference day 2008-08-24, plotted in black, as measured by the SIM instrument onboard SORCE. For comparison, we also show Planck distributions for 6500 K in red, 5770 K in blue, and 4500 K in green. The Planck distributions use Equation (D2) for a fixed temperature, with wavelength as independent variable, and transformed to spectral irradiance by multiplying by the factor $\alpha_s = \pi * \left(\frac{R_s}{AU}\right)^2 = 6.79426 * 10^{-5}$, with $R_s$ being the Sun's mean radius, and $AU$ the mean Earth-Sun distance, as in Equation (D3).

**Figure 2.** A zoom of Figure 1 for wavelength range 200 nm to 600 nm. The apparently irregular bumps in this plot, and in Figure 1, are due to well-known Frauenhofer lines in the solar spectrum, smoothed to the SIM instrument's bandpass, which varies from about 1 nm width near wavelength 200 nm, up to about 30 nm width near 1000 nm, then decreasing slightly. The width of a typical atomic Frauenhofer line is of order 1 Angstrom, or 0.1 nm, so many of the observed bumps are smoothed clusters of nearby lines. A few of the contributing atomic lines are indicated in the labels on the vertical dashed lines. For example, the green dashed line near 430 nm, is labelled CaFeg to indicate that lines of calcium, iron, and oxygen (g-band) are all included within the plotted bump in the black line. Effects of ionization thresholds are also seen, such as just above the CaII H and K lines near 400 nm, which has photon energies near 3.1 eV.

**Figure 3.** Temperature vs Wavelength on the reference day, 2008-08-24. The plot in **A)** shows the same wavelength range as in Figure 1. Plot **B)** is a zoom into the same short wavelength range as in Figure 2. As in Figure 2, several bumps are labelled with contributing atomic lines, such as the green dashed line near 430 nm, labelled CaFeg (calcium, iron, oxygen g-band). As in Figure 2, the rise due to the ionization threshold is evident near 400 nm, just above the CaII H and K lines. In both plots, temperature at each wavelength was computed using a *Mathematica* root-finding procedure to solve for *T* in Equation (D3), $SSI = \alpha_s B(\lambda, T)$, with *SSI* the observed value.

**Figure 4.** Time series of irradiance for all records of daily average data from the full 17 years of SIM data, version 27, downloaded from LISIRD. In this case we have chosen the $H_\alpha$ wavelength, 656.2 nm. In this plot, it is evident that there is a minimum of solar activity in mid 2008. We choose as a reference day 2008-08-24, and consider variations about this day to approximate the temperatures on all other days.

**Figure 5.** Time series of the Temperature calculated in *Wolfram Mathematica* for all records of solar spectral data with fixed wavelength, $H_\alpha$ = 656.20 nm, using Equation (D3). We term the root-finding solution of Equation (D3) the "exact" value of the temperature, to distinguish it from the two analytic approximations (linear and quadratic) described in appendices D and E, and from the two statistical "fit" approximations (also linear and quadratic) described in appendix G.

**Figure 6.** Linear analytic approximation of the temperature compared with the exact value (value obtained by root-finding solution of Equation (D3). The linear analytic approximation is given in Equation (D14).

**Figure 7. A)** The difference between the exact value of *T* and the value obtained with the linear analytic approximation. **B)** Delta linear approximation of the temperature using the Equation 3.1 and expressed in parts per million (ppm). The ***negative*** values in this figure indicate that the linear analytic approximation ***overestimates*** the exact value of the temperature.

**Figure 8.** Quadratic analytic approximation of *T* compared with the exact value. The quadratic approximation model is given in Equation (E3).

**Figure 9.** (A) The difference between the exact value of *T* and the value obtained with the quadratic analytic approximation. (B**)** Delta quadratic approximation of *T* using Equation 3.1 and expressed in part per million (ppm). The ***positive*** values in this figure indicate that the quadratic analytic approximation ***underestimates*** the exact value of the temperature.  Combining this result with that of Figure 7 shows that the exact value lies ***between*** the linear and quadratic analytic approximations.

**Figure 10.**  (A) Irradiance versus time during 2003-2008, for four chosen wavelengths, overlaid on a plot of temperature *T* versus wavelength on the reference day, and indicating the reference day temperatures at those same four chosen wavelengths. (B) Irradiance versus time during 2003-2020, for the shortest chosen wavelength, 285.48 nm. (C) Irradiance versus time during 2003-2020, for the second chosen wavelength, 619.4 nm. (D) Irradiance versus time during 2003-2020, for the third chosen wavelength, 855.93 nm. (E) Irradiance versus time during 2003-2020, for the longest chosen wavelength, 1547.09 nm.

**Figure 11.** For each wavelength there is an associated linear "fit" coefficient, calculated using Mathematica and R software. (A) The ratio obtained from the division of the linear coefficients between the respective wavelengths. (B) The differences between the ratios. (C) Comparison between the analytic ratio ***a′*** and the value obtained with the polynomial interpolation in the same range of wavelengths, as in Equations (I1) and (I2).

**Figure 12.**  Three curves where: Diff is the difference between analytic linear coefficient ***a′*** obtained with Equations (G4) or (D13), and the fit linear coefficient ***a*** obtained with Equation (G1), for selected wavelengths, using either the complete data from 2003 to 2020, or only early data in range R1 to compute ***a*R1** or only late data in range R2 to compute ***a*R2**. In brief, R1=2003–2010, while R2=2011–2020.  Note that ***a*R1** and ***a*R2** values lie on either side of the overall difference value of ***a***, for each wavelength.

**Figure 13.**  The relative error between analytic linear coefficient ***a′*** and the fit linear coefficient ***a***, where fit coefficient ***a*** is calculated in the same three ways as in Figure 12, namely using the full available time period 2003–2020, or R1=2003–2010, or R2=2011–2020.

## Figures

**Figure 1. Irradiance as a function of wavelength on the reference day**

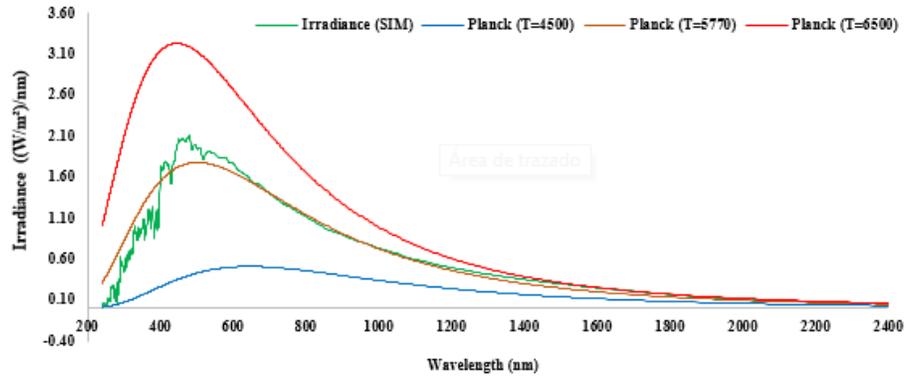

**Figure 2. As in Figure 1, wavelengths 240 – 660 nm, some groups of spectral lines labelled**

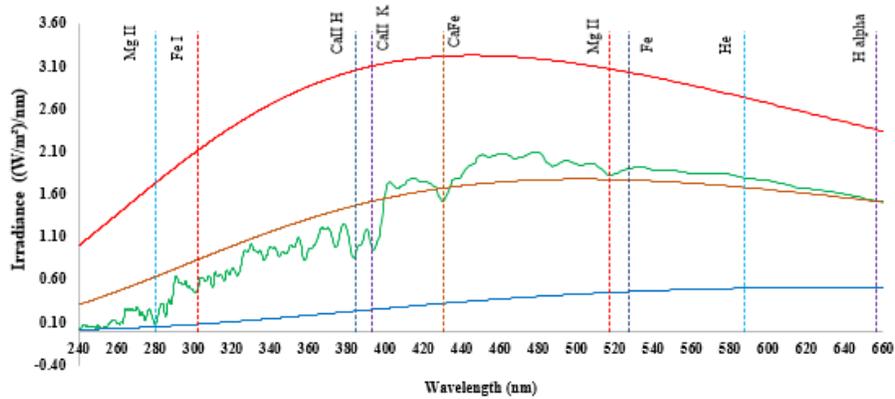

**Figure 3. The temperature on reference day**

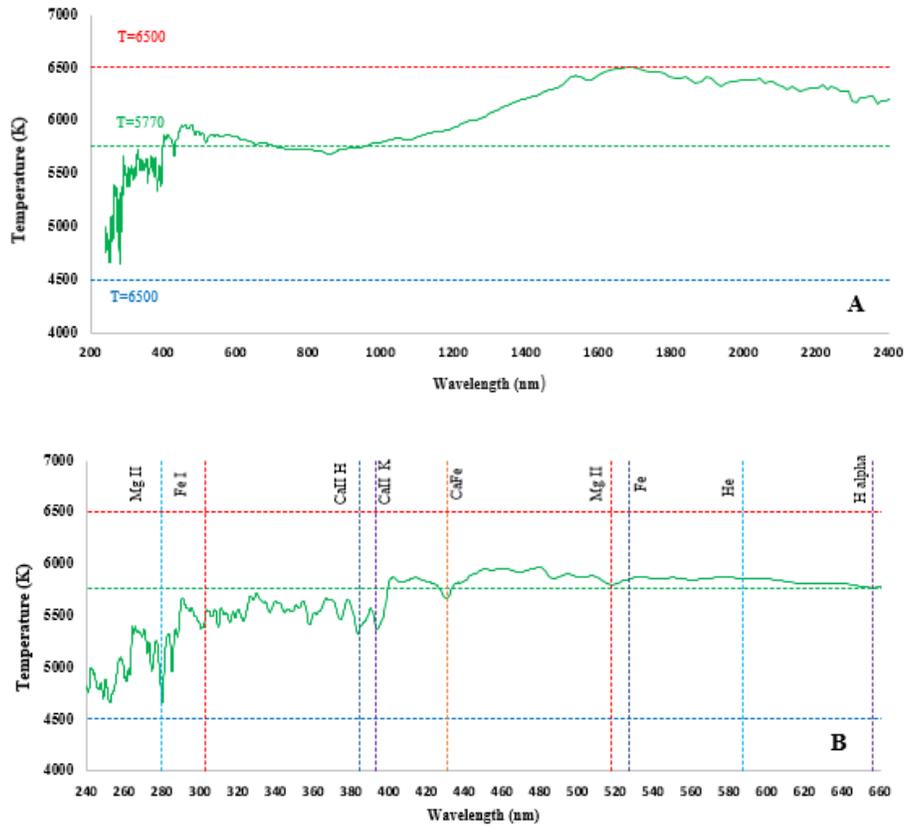

Figure 4. Solar spectral irradiance of $H_\alpha$ (656.20 nm)

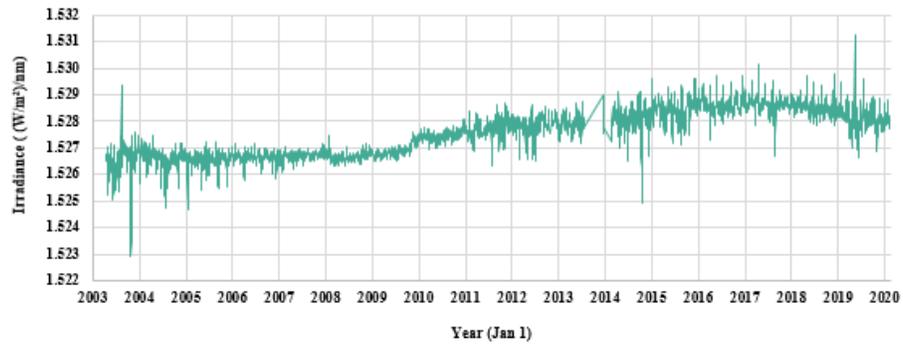

Figure 5. The temperature of $H_\alpha$ (656.20 nm)

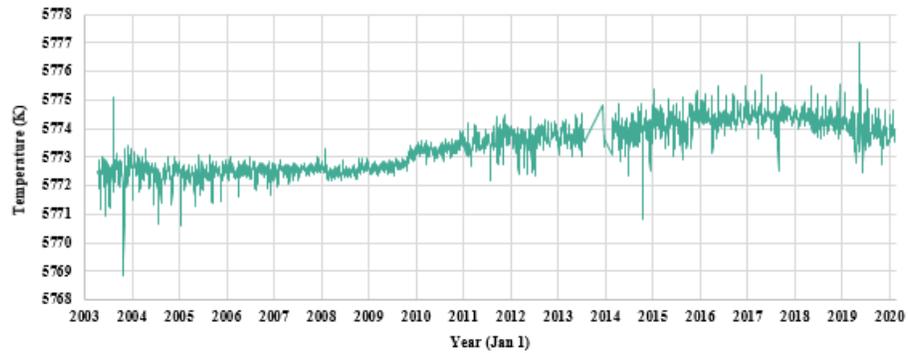

**Figure 6. Linear Approximation and Temperature of $H_\alpha$ (656.20 nm)**

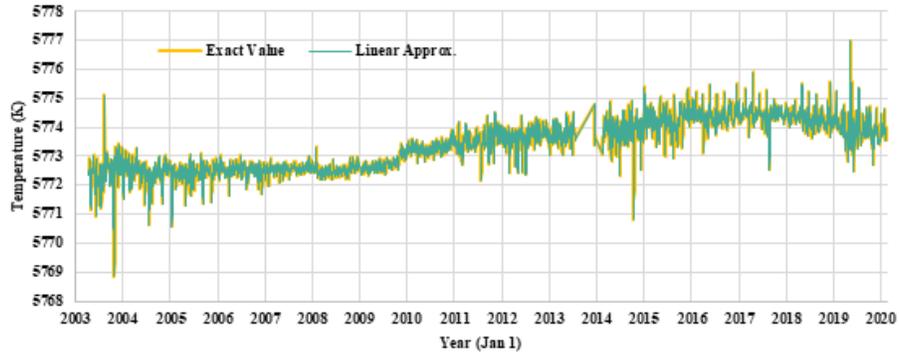

**Figure 7. Difference and Delta of the Linear Approximation of the Temperature $H_\alpha$ (656.20 nm)**

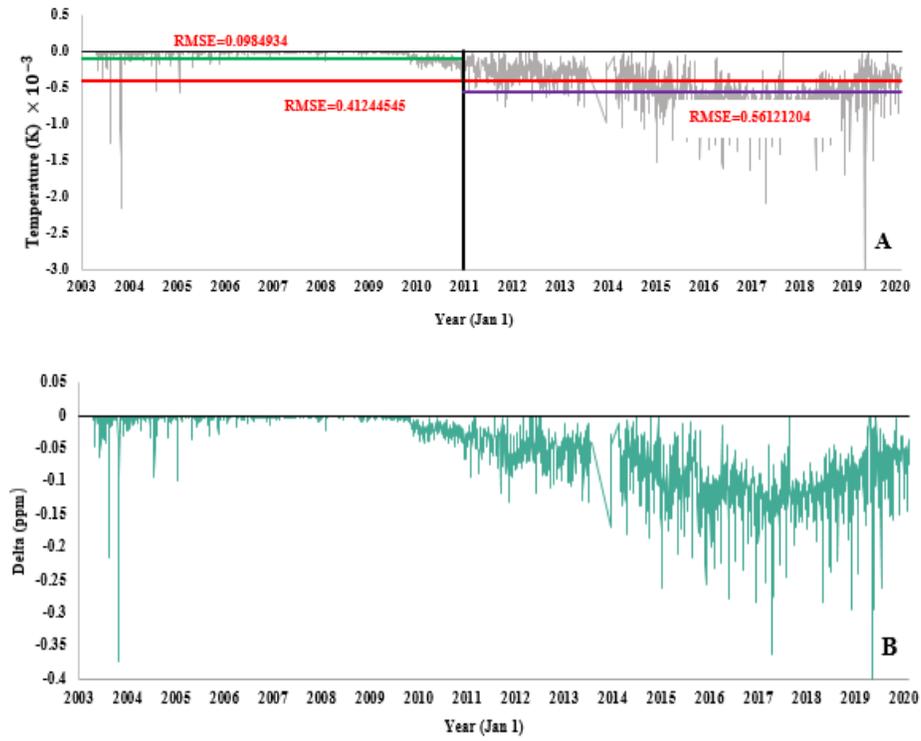

**Figure 8. Quadratic Approximation and Temperature $H_\alpha$ (656.20 nm)**

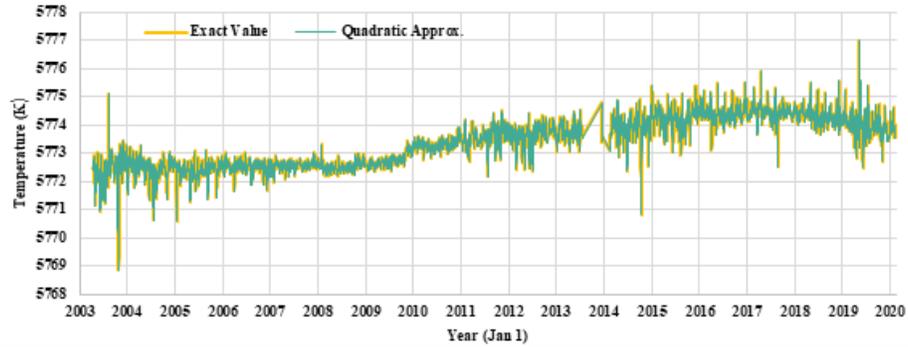

# Figure 9. Difference and Delta of the Quadratic Approximation of the Temperature $H_\alpha$ (656.2 nm)

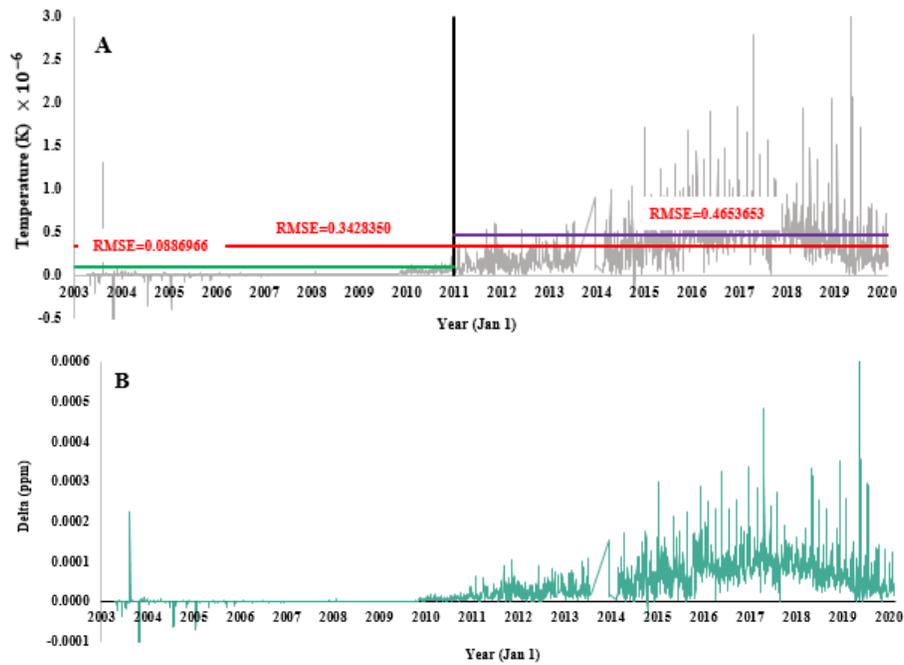

**Figure 10. Spectral Variability Nomogram**

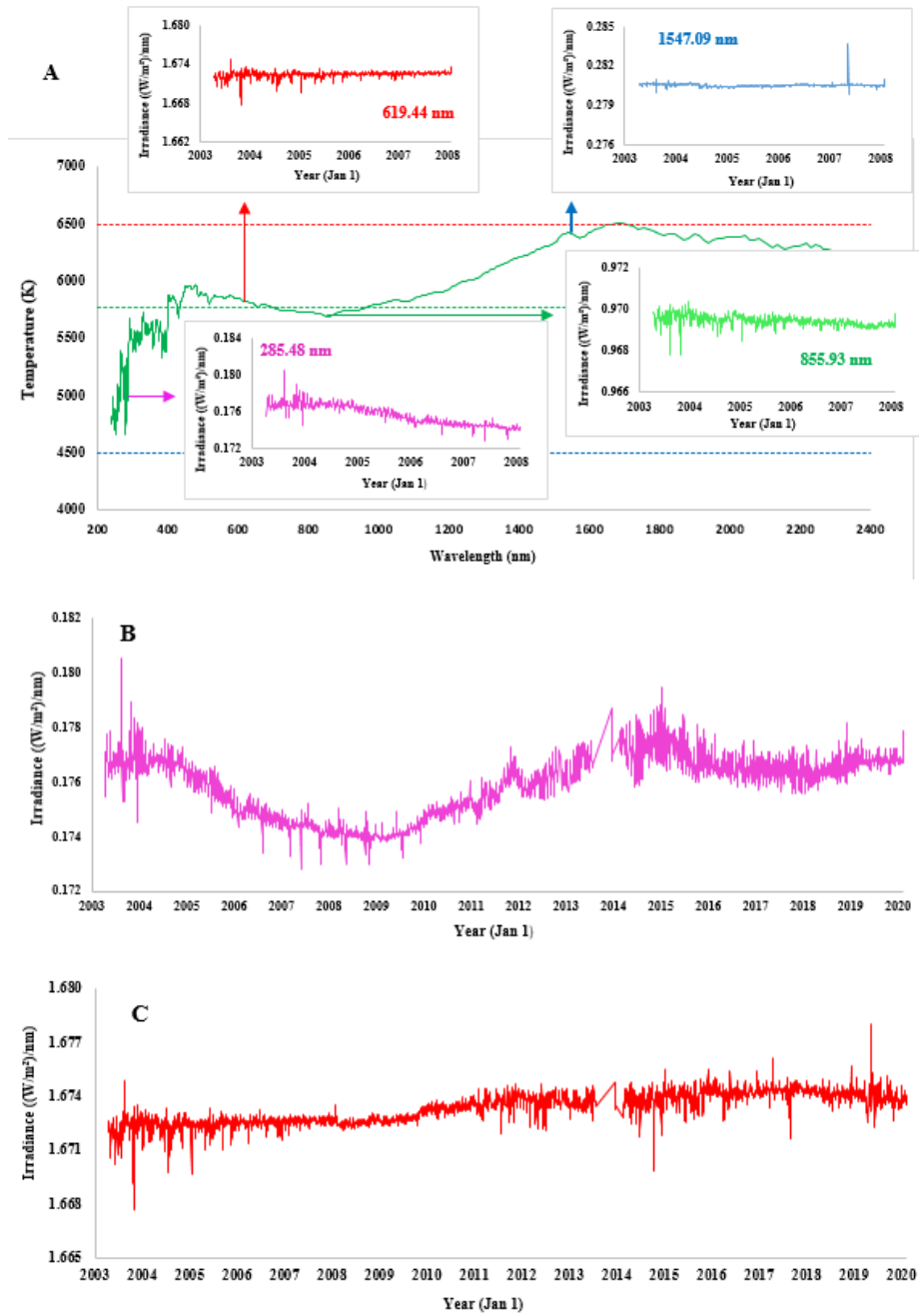

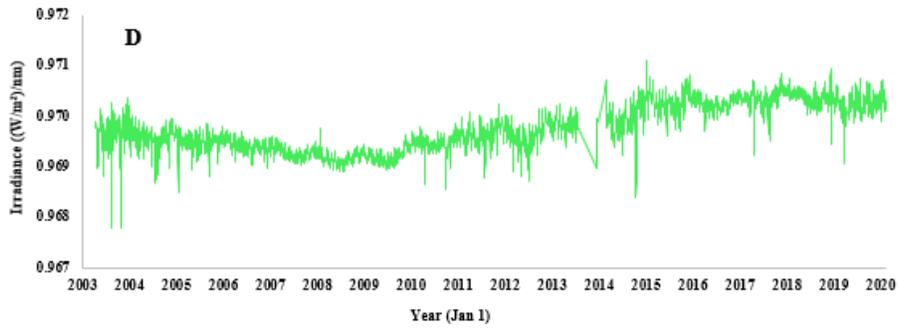

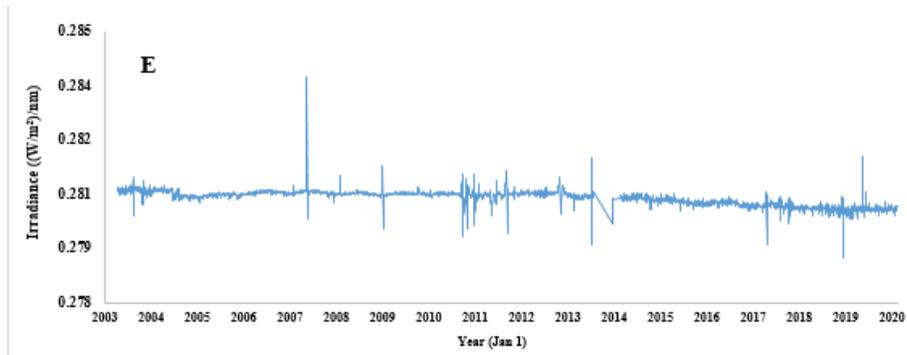

**Figure 11: Ratio of the linear coefficient and the interpolation model**

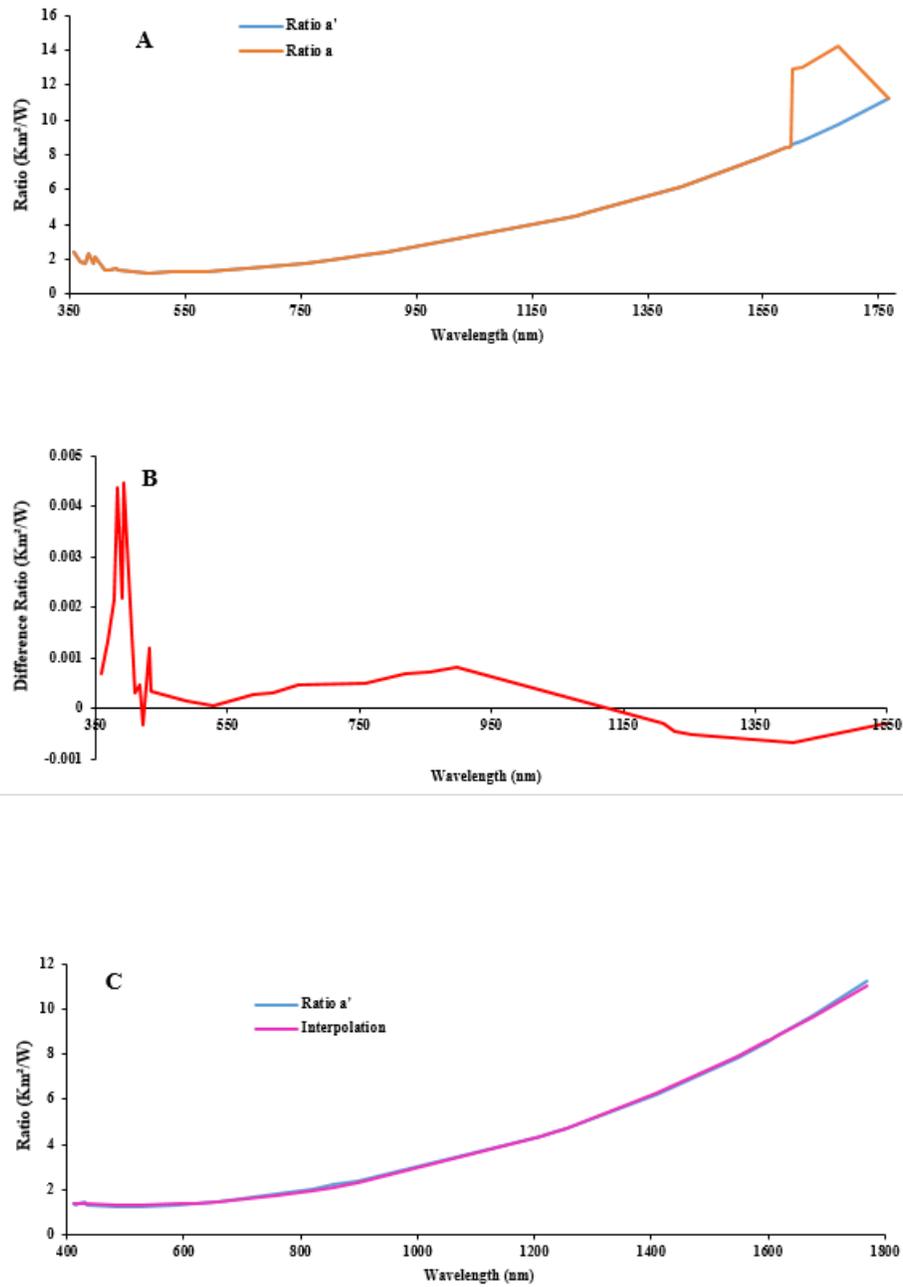

**Figure 12: Difference of constants $a'$ and $a$ for each range of time**

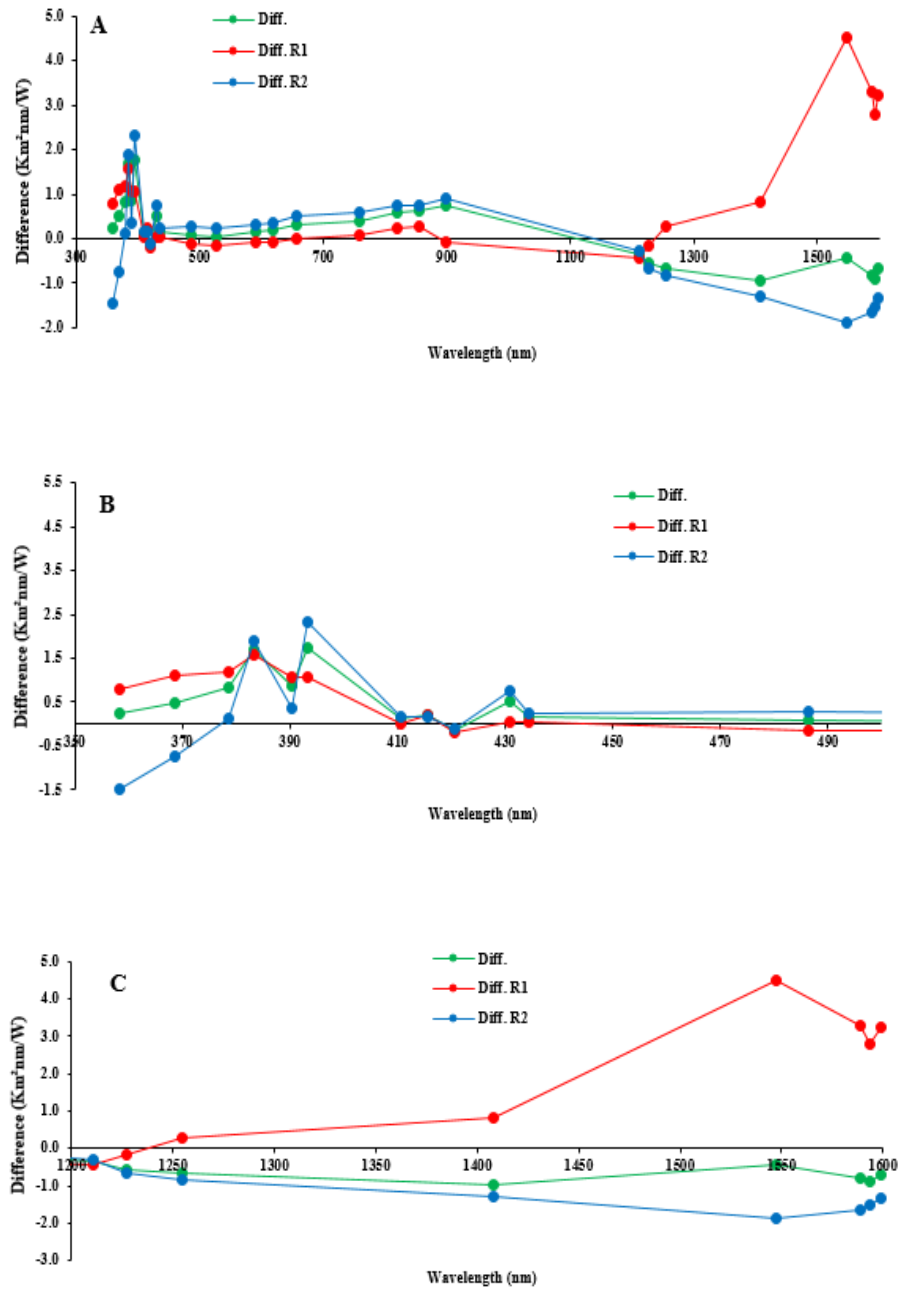

**Figure 13: Relative error**

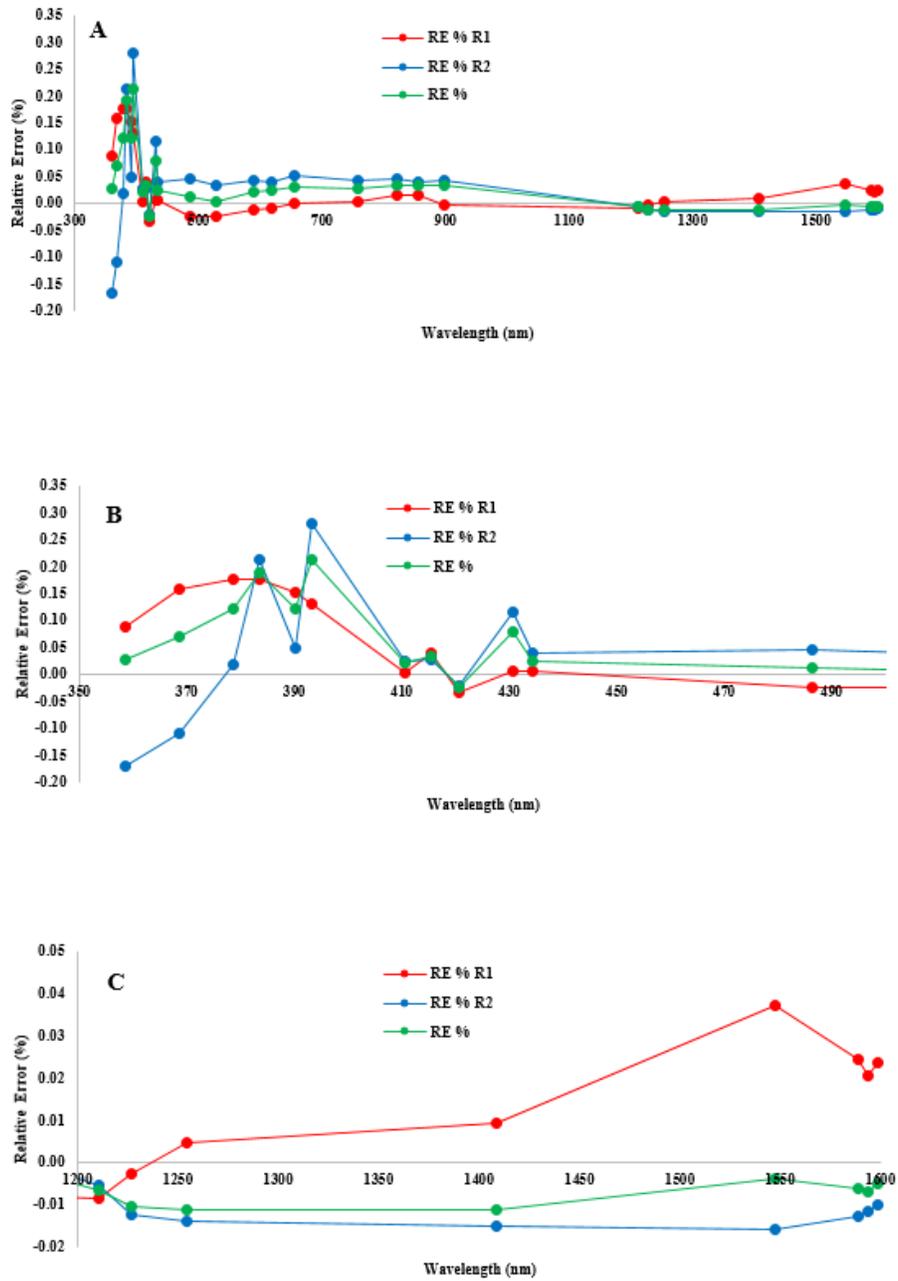

## Table Captions

**Table 1:** Root mean square error (RMSE) and the Mean error (ME) of the temperature values obtained with the different models compared with exact Temperature values, for 2003 to 2020, and 285.48 nm.

**Table 2:** RMSE and ME of the temperature values obtained with the different models compared with the exact Temperature values, over full range 2003 to 2020, for wavelength 656.20 nm ($H_\alpha$).

**Table 3:** RMSE and ME of the temperature values obtained with the different models compared with the exact Temperature values from 2003 to 2020 for the wavelength 855.93 nm.

**Table 4:** RMSE and ME of the temperature values obtained with the different models compared with the exact Temperature values from 2003 to 2020 for the wavelength 1547.09 nm.

**Table 5:** In order to approximate the values of the temperature *T*, for a chosen set of wavelengths, on any chosen date, using Equations (D14) and (E3), the values of *T* and *SSI* on the reference day ($T_o$, $SSI_o$) are needed, as are the values of *SSI* on the day that the approximate values of *T* are desired, for instance 2011-10-10. Also displayed in this table are the "exact" temperature values calculated by Equation (D2), and verified by solving Equation (D3) by root-finding in *Wolfram Mathematica* for each wavelength, on the date of our example, 2011-10-10.

**Table 6:** Values of linear ($K/((W/m^2)/nm)$) and quadratic ($K/((W/m^2)/nm)^2$) coefficients are shown, as given in Equations (D13) and (E2) evaluated for the data of the reference day 2008-08-24, taken from Table 5.

**Table 7:** Temperature values on 2011-10-10 for four wavelengths using the linear approximation of Equation (G4), compared with the exact value obtained using Equation (D2), and also (as a check) by solving Equation (D3) in *Wolfram Mathematica*.

**Table 8:** Temperature values of 2011-10-10 for four wavelengths using the quadratic approximation model of Equation (G5), compared with the exact value obtained using Equation (D2), and checked by solving Equation (D3) in *Wolfram Mathematica*.

**Table 9:** Parameters "a" and "b" of the linear fit model (G1) for each wavelength, which were obtained with *R* software, necessary to estimate the temperature on 2011-10-10.

**Table 10:** Parameters "A", "B" and "C" of the quadratic fit model (G2) are shown for each wavelength. Here we can see that the quadratic adjustment improves to a certain extent the value of the estimated temperature of 2011-10-10.

**Table 11:** The constant $a'$ calculated using Equation (G4) or (D13), and the constant $a$ calculated using Equation (G1) of the linear fit model, using *R* software with data from 2003 to 2020.

**Table 12:** Values of the linear fit constant $a$ calculated using two ranges of dates, R1 and R2: $a$ R1 is the linear coefficient calculated with *R* software using data from 2003 to 2010, while $a$ R2 is calculated using data from 2011 to 2020. The relative errors are also obtained when making the comparison with $a'$.

**Table 13:** Values of the quotients $a'$ and $a$ for different wavelengths and the values of $a'$ obtained using the polynomial interpolation of Equations (I1) and (I2).

## Tables

**Table 1:** RMSE and Mean Error values for 285.48 nm

| Model | RMSE (K) | Mean Error (K) |
|---|---|---|
| Linear analytic | 0.0416885145 | -0.0319543621 |
| Quadratic analytic | 0.0004726288 | 0.000313800 |
| Linear fit | 0.0081867338 | -0.0000000031 |
| Quadratic fit | 0.0000457226 | -0.0000000005 |

**Table 2:** RMSE and Mean Error values for $H_\alpha$ 656.20 nm

| Model | RMSE (K) | Mean Error (K) |
|---|---|---|
| Linear analytic | 0.0004124455 | -0.0002771424 |
| Quadratic analytic | 0.0000003428 | 0.0000001945 |
| Linear fit | 0.0001449298 | 0.0000000144 |
| Quadratic fit | 0.0000001221 | 0.0000000304 |

**Table 3:** RMSE and Mean Error values for 855.93 nm

| Model | RMSE (K) | Mean Error (K) |
|---|---|---|
| Linear analytic | 0.0004235579 | -0.0003016674 |
| Quadratic analytic | 0.0000003408 | 0.0000002162 |
| Linear fit | 0.0000857852 | 0.0000000079 |
| Quadratic fit | 0.0000000438 | 0.0000000091 |

**Table 4:** RMSE and Mean Error values for 1547.09 nm

| Model | RMSE (K) | Mean Error (K) |
|---|---|---|
| Linear analytic | 0.0007927592 | -0.0002137115 |
| Quadratic analytic | 0.0000045085 | -0.0000000978 |
| Linear fit | 0.0007581064 | 0.0000000087 |
| Quadratic fit | 0.0000022569 | 0.0000000108 |

**Table 5:** data of solar spectral irradiance in the reference day and of other day, 2011-10-10.

| Wavelength (nm) | $SSI_o$ ($\frac{W}{m^2}$/nm) | $T_o$ (K) | SSI ($\frac{W}{m^2}$/nm) | Exact value of T (K) |
|---|---|---|---|---|
| 285.48 | 0.1739754 | 4985.44659842 | 0.1759321 | 4990.9681473 |
| 656.20 | 1.526558 | 5772.41067100 | 1.527622 | 5773.4459772 |
| 855.93 | 0.9690168 | 5688.34171545 | 0.9696425 | 5689.5197810 |
| 1547.09 | 0.2805222 | 6417.67574425 | 0.2805273 | 6417.7373565 |

**Table 6:** linear and quadratic analytic coefficients calculated for the reference day.

| Wavelength (nm) | Quadratic coefficient $\left(\frac{d^2T}{dSSI^2}\right)_o$ | Linear coefficient $\left(\frac{dT}{dSSI}\right)_o$ |
|---|---|---|
| 285.48 | -13070.296 | 2834.568 |
| 656.20 | -323.644 | 973.204 |
| 855.93 | -797.737 | 1883.046 |
| 1547.09 | -7693.756 | 12080.859 |

**Table 7:** Temperature values obtained with the linear analytic approximation model for the example day 2011-10-10.

| Wavelength (nm) | Constant $a'$ (K/($\frac{W}{m^2}$/nm)) | Constant $b'$ (K) | Estimated T (K) | Difference (K) |
|---|---|---|---|---|
| 285.48 | 2834.568 | 4492.301 | 4990.9929982 | -0.0248510 |
| 656.20 | 973.204 | 4286.758 | 5773.4461603 | -0.0001831 |
| 855.93 | 1883.046 | 3863.639 | 5689.5199371 | -0.0001561 |
| 1547.09 | 12080.859 | 3028.727 | 6417.7373566 | -0.0000001 |

**Table 8:** Temperature values obtained with the quadratic analytic approximation model for the example day 2011-10-10.

| Wavelength (nm) | Constant $C'$ $(K/(\frac{W}{m^2}/nm)^2)$ | Constant $A'$ $(K/(\frac{W}{m^2}/nm))$ | Constant $B'$ (K) | Estimated T (K) | Difference (K) |
|---|---|---|---|---|---|
| 285.48 | -6535.148147 | 5108.478341 | 4294.499239 | 4990.9679773 | 0.00016998 |
| 656.20 | -161.8219945 | 1467.265593 | 3909.651272 | 5773.4459771 | 0.00000008 |
| 855.93 | -398.8687194 | 2656.066599 | 3489.103717 | 5689.5197809 | 0.00000007 |
| 1547.09 | -3846.878 | 14239.12872 | 2726.005277 | 6417.7373565 | 0.00000000 |

**Table 9:** Constants *a* and *b* of the linear fit model, and the temperatures *T* estimated with the linear fit model, compared with the exact values given in Table 5.

| Wavelength (nm) | Constant $a$ $(K/(W/m^2/nm))$ | Constant b (K) | Estimated T (K) | Difference (K) |
|---|---|---|---|---|
| 285.48 | 2811.638 | 4496.303 | 4990.9602941 | 0.0078532 |
| 656.20 | 972.896 | 4287.228 | 5773.4458560 | 0.0001212 |
| 855.93 | 1882.429 | 3864.236 | 5689.5197041 | 0.0000769 |
| 1547.09 | 12081.303 | 3028.602 | 6417.7371987 | 0.0001579 |

**Table 10:** Constants *A, B, C* of the quadratic fit model, and the temperatures *T* estimated with the quadratic fit model, compared with the exact values taken from Table 5.

| Wavelength (nm) | Constant C $(K/(\frac{W}{m^2}/nm)^2)$ | Constant A $(K/(\frac{W}{m^2}/nm))$ | Constant B (K) | Estimated T (K) | Difference (K) |
|---|---|---|---|---|---|
| 285.48 | -6395.976 | 5059.831 | 4298.750 | 4990.9681295 | 0.0000178 |
| 656.20 | -161.712 | 1466.929 | 3909.908 | 5773.4459773 | 0.0000000 |
| 855.93 | -398.428 | 2655.212 | 3489.518 | 5689.5197810 | 0.0000000 |
| 1547.09 | -3829.317 | 14229.287 | 2727.384 | 6417.7373568 | -0.0000003 |

**Table 11:** Constant $a'$ calculated with the linear analytic approximation and constant $a$ calculated with the linear fit model for certain wavelengths using data from 2003 to 2020.

| Wavelength (nm) | Constant $a'$ (K/ (W/m²/nm)) | Constant $a$ (K/ (W/m²/nm)) | Difference (K/ (W/m²/nm)) | Relative Error % |
|---|---|---|---|---|
| 358.48 | 873.388 | 873.145 | 0.243 | 0.03 |
| 368.52 | 684.319 | 683.826 | 0.494 | 0.07 |
| 378.65 | 671.005 | 670.186 | 0.819 | 0.12 |
| 383.35 | 879.895 | 878.216 | 1.679 | 0.19 |
| 390.26 | 700.309 | 699.454 | 0.855 | 0.12 |
| 393.36 | 823.076 | 821.323 | 1.752 | 0.21 |
| 410.70 | 560.720 | 560.594 | 0.126 | 0.02 |
| 415.60 | 554.559 | 554.369 | 0.190 | 0.03 |
| 420.68 | 566.168 | 566.304 | -0.135 | -0.02 |
| 430.77 | 632.893 | 632.381 | 0.511 | 0.08 |
| 434.31 | 583.383 | 583.232 | 0.151 | 0.03 |
| 486.32 | 597.197 | 597.124 | 0.073 | 0.01 |
| 527.52 | 655.268 | 655.235 | 0.033 | 0.01 |
| 589.35 | 772.610 | 772.446 | 0.164 | 0.02 |
| 619.44 | 855.672 | 855.470 | 0.202 | 0.02 |
| 656.20 | 973.204 | 972.896 | 0.308 | 0.03 |
| 758.01 | 1358.396 | 1358.008 | 0.388 | 0.03 |
| 820.85 | 1670.225 | 1669.646 | 0.579 | 0.03 |
| 855.93 | 1883.046 | 1882.429 | 0.616 | 0.03 |
| 897.83 | 2128.196 | 2127.466 | 0.730 | 0.03 |
| 1210.76 | 5263.981 | 5264.319 | -0.338 | -0.01 |
| 1227.20 | 5493.583 | 5494.151 | -0.568 | -0.01 |
| 1254.68 | 5910.298 | 5910.953 | -0.655 | -0.01 |
| 1408.28 | 8702.579 | 8703.534 | -0.955 | -0.01 |
| 1547.09 | 12080.859 | 12081.303 | -0.444 | 0.00 |
| 1588.65 | 13326.562 | 13327.369 | -0.807 | -0.01 |
| 1593.80 | 13477.502 | 13478.401 | -0.899 | -0.01 |
| 1598.95 | 13630.355 | 13631.038 | -0.683 | -0.01 |
| 1601.18 | 13701.537 | 20644.058 | -6942.520 | -50.67 |
| 1616.86 | 14180.873 | 21111.621 | -6930.748 | -48.87 |
| 1678.61 | 16258.404 | 23817.622 | -7559.217 | -46.49 |
| 1768.43 | 19787.619 | 19800.177 | -12.558 | -0.06 |

**Table 12:** Constant *a* **R1** calculated for certain wavelengths using data from 2003 to 2010 and constant *a* **R2** using data from 2011 to 2020. Also shown are relative errors RE, compared to the linear analytic model that uses coefficient *a′*.

| Wavelength (nm) | Constant a R1 | Constant a R2 | Diff. R1 | Diff. R2 | RE % R1 | RE % R2 |
|---|---|---|---|---|---|---|
| **358.48** | 872.602 | 874.854 | 0.787 | -1.466 | 0.09 | -0.17 |
| **368.52** | 683.224 | 685.061 | 1.095 | -0.742 | 0.16 | -0.11 |
| **378.65** | 669.821 | 670.869 | 1.183 | 0.135 | 0.18 | 0.02 |
| **383.35** | 878.333 | 878.005 | 1.562 | 1.890 | 0.18 | 0.21 |
| **390.26** | 699.234 | 699.953 | 1.076 | 0.356 | 0.15 | 0.05 |
| **393.36** | 822.001 | 820.758 | 1.074 | 2.318 | 0.13 | 0.28 |
| **410.70** | 560.697 | 560.571 | 0.024 | 0.149 | 0.00 | 0.03 |
| **415.60** | 554.337 | 554.397 | 0.222 | 0.162 | 0.04 | 0.03 |
| **420.68** | 566.357 | 566.278 | -0.189 | -0.109 | -0.03 | -0.02 |
| **430.77** | 632.841 | 632.157 | 0.052 | 0.736 | 0.01 | 0.12 |
| **434.31** | 583.339 | 583.147 | 0.044 | 0.237 | 0.01 | 0.04 |
| **486.32** | 597.333 | 596.913 | -0.135 | 0.285 | -0.02 | 0.05 |
| **527.52** | 655.425 | 655.043 | -0.156 | 0.226 | -0.02 | 0.03 |
| **589.35** | 772.700 | 772.284 | -0.090 | 0.326 | -0.01 | 0.04 |
| **619.44** | 855.733 | 855.319 | -0.061 | 0.353 | -0.01 | 0.04 |
| **656.20** | 973.209 | 972.702 | -0.005 | 0.503 | 0.00 | 0.05 |
| **758.01** | 1358.331 | 1357.818 | 0.064 | 0.578 | 0.00 | 0.04 |
| **820.85** | 1669.971 | 1669.470 | 0.254 | 0.755 | 0.02 | 0.05 |
| **855.93** | 1882.755 | 1882.294 | 0.291 | 0.751 | 0.02 | 0.04 |
| **897.83** | 2128.273 | 2127.304 | -0.076 | 0.892 | 0.00 | 0.04 |
| **1210.76** | 5264.415 | 5264.262 | -0.434 | -0.281 | -0.01 | -0.01 |
| **1227.20** | 5493.734 | 5494.265 | -0.151 | -0.682 | 0.00 | -0.01 |
| **1254.68** | 5910.018 | 5911.117 | 0.280 | -0.819 | 0.00 | -0.01 |
| **1408.28** | 8701.740 | 8703.884 | 0.838 | -1.305 | 0.01 | -0.02 |
| **1547.09** | 12076.345 | 12082.741 | 4.514 | -1.882 | 0.04 | -0.02 |
| **1588.65** | 13323.272 | 13328.223 | 3.291 | -1.661 | 0.02 | -0.01 |
| **1593.80** | 13474.719 | 13479.027 | 2.783 | -1.525 | 0.02 | -0.01 |
| **1598.95** | 13627.132 | 13631.683 | 3.223 | -1.328 | 0.02 | -0.01 |

**Table 13:** Constants $a'$, $a$ and the values of $a'$ calculated with the polynomial interpolation of Equations (I1) and (I2).

| Wavelength (nm) | Ratio a' (K/W/m²) | Ratio a (K/W/m²) | Diff (a'-a) (K/W/m²) | Interpolation Ratio (K/W/m²) |
|---|---|---|---|---|
| 410.70 | 1.3653 | 1.3650 | 0.0003 | 1.3747 |
| 415.60 | 1.3344 | 1.3339 | 0.0005 | 1.3694 |
| 420.68 | 1.3458 | 1.3462 | -0.0003 | 1.3642 |
| 430.77 | 1.4692 | 1.4680 | 0.0012 | 1.3548 |
| 434.31 | 1.3432 | 1.3429 | 0.0003 | 1.3518 |
| 486.32 | 1.2280 | 1.2278 | 0.0002 | 1.3251 |
| 527.52 | 1.2422 | 1.2421 | 0.0001 | 1.3272 |
| 589.35 | 1.3110 | 1.3107 | 0.0003 | 1.3688 |
| 619.44 | 1.3814 | 1.3810 | 0.0003 | 1.4058 |
| 656.20 | 1.4831 | 1.4826 | 0.0005 | 1.4658 |
| 758.01 | 1.7921 | 1.7915 | 0.0005 | 1.7173 |
| 820.85 | 2.0348 | 2.0340 | 0.0007 | 1.9351 |
| 855.93 | 2.2000 | 2.1993 | 0.0007 | 2.0774 |
| 897.83 | 2.3704 | 2.3696 | 0.0008 | 2.2669 |
| 1210.76 | 4.3477 | 4.3479 | -0.0003 | 4.3532 |
| 1227.20 | 4.4765 | 4.4770 | -0.0005 | 4.4955 |
| 1254.68 | 4.7106 | 4.7111 | -0.0005 | 4.7407 |
| 1408.28 | 6.1796 | 6.1803 | -0.0007 | 6.2794 |
| 1547.09 | 7.8088 | 7.8090 | -0.0003 | 7.9152 |
| 1588.65 | 8.3886 | 8.3891 | -0.0005 | 8.4503 |
| 1593.80 | 8.4562 | 8.4568 | -0.0006 | 8.5180 |
| 1598.95 | 8.5246 | 8.5250 | -0.0004 | 8.5861 |
| 1601.18 | 8.5954 | 12.8930 | -4.2977 | 8.6157 |
| 1616.86 | 8.8100 | 13.0572 | -4.2472 | 8.8254 |
| 1678.61 | 9.7296 | 14.1889 | -4.4593 | 9.6800 |
| 1768.43 | 11.2408 | 11.1965 | 0.0443 | 11.0054 |